\documentclass[aps,prd,10pt,preprintnumbers,nofootinbib,superscriptaddress,notitlepage]{revtex4-2}

\usepackage{amsmath}
\usepackage{amssymb}
\usepackage{latexsym}
\usepackage{amsfonts}
\usepackage{epsfig}
\usepackage{graphicx}
\usepackage{wasysym}
\usepackage{ulem}
\usepackage{xcolor}
\usepackage{subfigure}
\usepackage{hyperref}

\allowdisplaybreaks
\newcommand{\Eqref}[1]{Eq.~\eqref{#1}}

\begin{document}

\title{Impact of background field localization on vacuum polarization effects}

\author{Carl Marmier}\email{carl.marmier@ens-paris-saclay.fr}
\affiliation{Universit\'{e} Paris-Saclay, ENS Paris-Saclay, DER de Physique, 91190, Gif-sur-Yvette, France.}
\affiliation{Helmholtz Institute Jena, Fr\"obelstieg 3, 07743 Jena, Germany}
\affiliation{GSI Helmholtzzentrum für Schwerionenforschung GmbH, Planckstra\ss e 1, 64291 Darmstadt, Germany}
\affiliation{Theoretisch-Physikalisches Institut, Abbe Center of Photonics, Friedrich-Schiller-Universit\"at Jena, Max-Wien-Platz 1, 07743 Jena, Germany}

\author{Nico Seegert}
\author{Felix Karbstein}\email{felix.karbstein@uni-jena.de}
\affiliation{Helmholtz Institute Jena, Fr\"obelstieg 3, 07743 Jena, Germany}
\affiliation{GSI Helmholtzzentrum für Schwerionenforschung GmbH, Planckstra\ss e 1, 64291 Darmstadt, Germany}
\affiliation{Theoretisch-Physikalisches Institut, Abbe Center of Photonics, Friedrich-Schiller-Universit\"at Jena, Max-Wien-Platz 1, 07743 Jena, Germany}

\begin{abstract}
We aim at insights about how localization of the background field impacts nonlinear quantum vacuum signatures probed by photons in purely magnetic, electric and crossed fields.
The starting point of our study are the one-loop results for the Heisenberg-Euler effective Lagrangian and the photon polarization tensor in quantum electrodynamics (QED) evaluated in a uniform constant electromagnetic field. 
As is well known and often employed, especially in the weak-field limit, within certain restrictions these results also allow for the reliable analysis of vacuum polarization effects in slowly varying background fields. 
Here, our main interest is in manifestly non-perturbative effects. To this end, we make use of the fact that for the particular case of background field inhomogeneities of Lorentzian shape with $0\leq d\leq3$ inhomogeneous directions analytical insights are possible. 
We study the scaling of conventional nonlinear QED signatures, such as 
probe-photon polarization flip and probe-photon induced electron-positron pair production, with relevant parameters.
Special attention is put on the $d$ dependence of the considered effects.
\end{abstract}

\date{\today}

\maketitle

\section{Introduction}

Charged particle-antiparticle fluctuations give rise to effective interactions between electromagnetic fields.
Within the Standard Model of particle physics these are mainly governed by quantum electrodynamics (QED), i.e., are driven by virtual electrons and positrons.
The fluctuations in particular affect the propagation of probe photons (four-potential $a^\mu$) sent through strong macroscopic electromagnetic fields (field strength tensor $F\equiv F^{\mu\nu}$) via the photon polarization tensor $\Pi^{\mu\nu}$. The latter encodes vacuum fluctuation mediated corrections to photon propagation at linear order in the fine-structure constant $\alpha=e^2/(4\pi)$ and beyond; $e>0$ denotes the elementary charge and we use Heaviside-Lorentz units with $c=\hbar=\epsilon_0=1$. 
In position space, the effective action at quadratic order in $a^\mu$ relevant for the study of probe photon propagation effects can be expressed as \cite{Dittrich:1985yb,Dittrich:2000zu}
\begin{equation}
    \Gamma[a]\big|_{\sim a^2}=-\frac{1}{4}\int{\rm d}^4x\, f_{\mu\nu}(x)f^{\mu\nu}(x)-\frac{1}{2}\int{\rm d}^4x'\int{\rm d}^4x\, a_\mu(x')\,\Pi^{\mu\nu}(x',x|F)\,a_\nu(x)\,, \label{eq:Gamma2}
\end{equation}
with probe field strength tensor $f^{\mu\nu}(x)=\partial^\mu a^\nu-\partial^\nu a^\mu$. Here, the polarization tensor specifically mediates an -- in general non-local -- interaction between in- and outgoing probe photons. Its last argument indicates its dependence on the background field $F$.
In the low-energy regime we have $\alpha\simeq1/137\ll1$, suggesting that the one-loop contribution linear in $\alpha$ constitutes the dominant fluctuation-induced correction; see also the reviews \cite{DiPiazza:2011tq,Battesti:2012hf,King:2015tba,Karbstein:2019oej,Fedotov:2022ely} and references therein.

Currently, only one-loop results for the photon polarization tensor in the presence of an external electromagnetic field $F$ are known explicitly, namely in uniform constant \cite{Batalin:1971au} and plane-wave \cite{Baier:1975ff,Becker:1974en} backgrounds.
In momentum space it is conventionally expressed as $\Pi^{\mu\nu}(k',k)\equiv\Pi^{\mu\nu}\bigl(k',k|F\bigr)$, where $k'\equiv k'^\mu$ and $k\equiv k^\mu$ denote the four-momenta of the in and out going photon, respectively; we use all-incoming conventions, i.e., four-momenta in the argument of $\Pi^{\mu\nu}(k',k)$ coming with the same sign are formally considered as incoming. We emphasize that $\Pi^{\mu\nu}(k',k|F)$ does not feature any explicit $x$ dependence via $F=F(x)$ in inhomogeneous fields because its derivation involves an integration over $x$.
The vacuum-fluctuation-induced effective couplings to $F(x)$ generically break the translational invariance of the vacuum and thereby result in a finite overlap between in and out states with $k'^\mu\neq k^\mu$.
Moreover, these may change the polarization properties of probe light because for two polarization vectors $\epsilon^\mu(k)$, $\epsilon_\perp^\mu(k)$ fulfilling $\epsilon_\perp^{*\mu}(k)\epsilon_\mu(k)=0$ the forward amplitude $\epsilon_\perp^{*\mu}(k)\Pi_{\mu\nu}(-k,k)\epsilon^\nu(k)$ can be non-vanishing.

Aside from the specific background field configurations mentioned above, analytical insights into the one-loop polarization tensor are possible (i) for slowly varying background fields of generic spatio-temporal structure, and photon momenta that are much smaller than the electron mass $m\simeq511\,{\rm keV}$, i.e. $\{|k'^\mu|,|k^\mu|\}\ll m$ for each component labeled by $\mu$.
Similarly, slowly varying fields are characterized by typical frequency scales of variation $\upsilon$ fulfilling $|\upsilon|\ll m$.
We emphasize that these encompass the subcategory of weakly localized fields, under which we understand fields that are slowly varying and -- at least in one direction $x_i$ -- confined to a finite space-time interval, in the sense that the field vanishes for $x_i\to\pm\infty$. 
The corresponding result for the photon polarization tensor \cite{Karbstein:2015cpa} can be readily extracted from the one-loop Heisenberg-Euler effective action in constant fields \cite{Heisenberg:1935qt,Weisskopf:1936hya,Schwinger:1951nm}; for reviews see Refs.~\cite{Dunne:2004nc,Dunne:2012vv}.
It is correct up to quadratic order in $k'^\mu\sim k^\mu$ and at zeroth-order in $\upsilon$.
Especially for magnetic and electric fields this result allows for controlled insights into both perturbative and manifestly non-perturbative parameter regimes in the coupling to the background field $F$.
Moreover, (ii) in the special case where the in an out momentum components equal each other, the photon polarization tensor can be reliably analyzed for slowly varying fields and arbitrary large values of the momentum transfer.
The restriction to $k'^\mu=k^\mu$ arises from the fact that it is impossible to recover the full momentum structure of $\Pi^{\mu\nu}\bigl(k',k|F\bigr)$ by adopting a slowly varying field approximation to the uniform constant field result $\Pi^{\mu\nu}(k',k|F={\rm const}.)\sim(2\pi)^4\delta^{(4)}(k'+k)$, with Dirac delta function $\delta^{(4)}(\cdot)$ in four space-time dimensions. See Refs.~\cite{Gies:2013yxa,Gies:2014wsa} and also the important clarification in Ref.~\cite{Karbstein:2015qwa}.
This approach (ii) is particularly relevant for the study of the polarization tensor in slowly varying inhomogeneous crossed fields fulfilling $|\vec{B}|=|\vec{E}|$ and $\vec{B}\cdot\vec{E}=0$ because in this case strategy (i) only provides access to the perturbative weak-field result at quadratic order in $F$ \cite{Karbstein:2015cpa}.
The reason for this is that the scalar field invariants ${\cal F}=F_{\mu\nu}F^{\mu\nu}/4$ and ${\cal G}=F_{\mu\nu}{}^\star\!F^{\mu\nu}/4$, with dual field strength tensor ${}^\star\!F^{\mu\nu}=\epsilon^{\mu\nu\rho\sigma}F_{\rho\sigma}/2$, vanish identically in crossed fields, such that higher powers in $F$ necessarily come with higher powers of $k'^\mu\sim k^\mu$.

In this work, we aim at analyzing how the localization of the background field influences the scaling of the photon polarization tensor with various parameters relative to the well-studied constant-field results. See also Ref.~\cite{Heinzl:2010vg} for a study of the impact of a finite pulse duration in stimulated laser pair production, and Ref.~\cite{Dinu:2014tsa} for a study of finite size effects on light-by-light scattering in the low-energy, weak-field regime.
To this end, we focus on parameter regimes where reliable approximate results can be obtained via (i) and (ii).
To be specific, we limit our analysis to background field inhomogeneities of Lorentzian shape with $0\leq d\leq3$ inhomogeneous directions, for which closed-form calculations are possible \cite{Gies:2013yxa,Karbstein:2017pbf}; see \Eqref{eq:calE} below for the explicit expression. 
We are convinced that such a study is very relevant, because it can provide us with important new insights in phenomena that are inaccessible with constant-field and plane-wave studies.
This becomes particularly evident for electromagnetic fields reaching near or above critical peak field strengths: especially in infinitely extended, constant electromagnetic fields, the regimes of perturbative weak and non-perturbative strong fields are conventionally clearly separated and amount to two complementary limits.
Conversely, probe light sent through localized inhomogeneous field configurations reaching non-perturbative peak field strengths inevitably experience both perturbative and non-perturbative field strengths.
Hence, we expect significant modifications in the strong-field scaling of quantum vacuum phenomena affecting the propagation of probe light.

Our article is organized as follows: In Sec.~\ref{sec:poltensor} we detail our derivation of the photon polarization tensor in the presence of a Lorentzian field inhomogeneity for the cases (i) and (ii).
In this context, we also extract approximate results in various limits.
Section~\ref{subsec:B+E} focuses on either a purely magnetic or electric field background, and Sec.~\ref{subsec:crossed} on a crossed-field configuration.
Subsequently, in Sec.~\ref{sec:phys}
we discuss the consequences of our findings on physical signatures of quantum vacuum nonlinearity that are -- at least in principle -- accessible in experiment.
Finally, we end with conclusions and a outlook in Sec.~\ref{sec:concls}.
Our metric convention is $g^{\mu\nu}={\rm diag}(-1,1,1,1)$.

\section{Photon polarization tensor}\label{sec:poltensor}

Our goal is to obtain analytical insights into the photon polarization tensor $\Pi^{\mu\nu}\bigl(k,k'|F(x)\bigr)$ in weakly localized background fields characterized by a Lorentzian amplitude profile
\begin{equation}
    {\cal E}(x)=\frac{{\cal E}_0}{1+\sum_{i=1}^d\bigl(\frac{2x_i}{w_i}\bigr)^2}\geq0\,,\quad\text{with}\quad 0\leq d\leq3\,.
    \label{eq:calE}
\end{equation}
Here, $d\in\mathbb{N}_0$ counts the number of inhomogeneous directions, ${\cal E}_0$ is the peak field amplitude and $w_i$ is the full width at half maximum (FWHM) in $i$ direction. For weakly localized background fields as considered throughout this work, by definition we have $w_i\gg\lambdabar_{\rm C}$, with reduced Compton wavelength of the electron $\lambdabar_{\rm C}=1/m$.
Clearly, in four space-time dimensions as considered here, a background field with $d$ inhomogeneous directions has $4-d$ homogeneous directions.
By construction, the uniform constant field result is retained for $d=0$. In the latter case the expression in \Eqref{eq:calE} reduces to ${\cal E}(x)={\cal E}_0$.

To be specific, in this work we only study the cases of either a purely magnetic $\vec{B}(x)={\cal E}(x)\,\hat{\vec{\kappa}}$ or electric $\vec{E}(x)={\cal E}(x)\,\hat{\vec{\kappa}}$ field pointing in a fixed direction $\hat{\vec{\kappa}}$, and a so-called crossed field characterized by perpendicular electric an magnetic fields featuring the same amplitude profile ${\cal E}(x)$ such that $\hat{\vec{\kappa}}=(\vec{E}\times\vec{B})/{\cal E}^2$ amounts to a globally fixed direction, respectively. Here, $\hat{\vec{\kappa}}$ denotes a unit vector.
For the purely magnetic or electric field case we exclusively focus on the regime of low momentum transfers and work out the exact expression for the polarization tensor at ${\cal O}(k^2)$ accessible via (i).
On the other hand, for the crossed-field configuration we resort to (ii).
However, for completeness we note that in the 'Tsai and Erber' regime \cite{Tsai:1974fa,Tsai:1975iz,Karbstein:2013ufa} characterized by a purely magnetic/electric field in conjunction with large-momentum probe photons the respective polarization tensor effectively reduces to the crossed field one, which directly implies that our crossed-field study is also relevant there.

\subsection{Magnetic or electric field}\label{subsec:B+E}

For electromagnetic fields fulfilling ${\cal G}=F_{\mu\nu}{}^\star\!F^{\mu\nu}/4=-\vec{E}\cdot\vec{B}=0$, encompassing the cases of purely magnetic or electric fields, the photon polarization tensor in momentum space extracted from the constant-field result of the one-loop Heisenberg-Euler effective action \cite{Heisenberg:1935qt,Weisskopf:1936hya,Schwinger:1951nm} can be compactly represented as \cite{Karbstein:2015cpa}
\begin{equation}
 \Pi^{\mu\nu}(-k',k)
 =\int{\rm d}^4x\,{\rm e}^{{\rm i}(k-k')x} \biggl[
 \bigl((k'k)g^{\mu\nu} - k'^\mu k^\nu \bigr)\pi_T 
 + \frac{(k'F)^\mu  (kF)^\nu}{2{\cal F}}\,\pi_{FF}
 + \frac{(k'{}^\star\!F)^\mu (k{}^\star\!F)^\nu}{2{\cal F}}\,\pi_{{}^\star\!F{}^\star\!F} 
 \biggr], \label{eq:PiF}
\end{equation}
where $2{\cal F}=F_{\mu\nu}F^{\mu\nu}/2=(\vec{B}^2-\vec{E}^2)$ and the scalar functions are given by
\begin{align}
    \pi_T&=-\frac{\alpha}{2\pi}\int_0^\infty\frac{{\rm d}s}{s}\,{\rm e}^{-\frac{m^2}{e\sqrt{2{\cal F}}}s}\biggl(\frac{1}{\sinh^2 s}-\frac{\coth s}{s}+\frac{2}{3}\biggr)\,,\label{eq:pi_T} \\
    \pi_{FF}&=-\frac{\alpha}{2\pi}\int_0^\infty\frac{{\rm d}s}{s}\,{\rm e}^{-\frac{m^2}{e\sqrt{2{\cal F}}}s}\biggl(\frac{1-2s\coth s}{\sinh^2 s}+\frac{\coth s}{s}\biggr)\,, \\
    \pi_{{}^\star\!F{}^\star\!F}&=-\frac{\alpha}{2\pi}\int_0^\infty\frac{{\rm d}s}{s}\,{\rm e}^{-\frac{m^2}{e\sqrt{2{\cal F}}}s}\biggl(\frac{1}{\sinh^2 s}-\frac{\coth s}{s}+\frac{2}{3}\,s\coth s\biggr)\,. \label{eq:pi_sFsF}
\end{align}
The parameter $s$ in Eqs.~\eqref{eq:pi_T}-\eqref{eq:pi_sFsF} is generically referred to as the propertime.
For later reference we note that in the special kinematic limit where $k'^\mu=k^\mu$ the tensor structure multiplying $\pi_T$ in \Eqref{eq:PiF} becomes equal to $(k^2g^{\mu\nu}-k^\mu k^\nu)\equiv k^2P_T^{\mu\nu}$, with transverse projector $P_T^{\mu\nu}$.
This is precisely the structure of the Maxwell term in \Eqref{eq:Gamma2} in momentum space because $-\frac{1}{4}\int{\rm d}^4x\,f_{\mu\nu}f^{\mu\nu}
 = -\frac{1}{2}\int\frac{{\rm d}^4k}{(2\pi)^4}\,a_\mu(k)\,k^2 P_T^{\mu\nu}a_\nu(k)$, and indicates that $\pi_T$ contains information about renormalization group properties of the theory.
Some additional clarifications are in order here:
Equation~\eqref{eq:PiF} neglects contributions scaling as $\sim k'k\bigl[{\cal O}\bigl((\frac{k}{m})^2\bigr)+{\cal O}\bigl((\frac{\upsilon}{m})^2\bigr)\bigr]$, where we count ${\cal O}(k')={\cal O}(k)$.
The square root in the exponential of Eqs.~\eqref{eq:pi_T}-\eqref{eq:pi_sFsF} is to be understood as $\sqrt{2{\cal F}}=\sqrt{2|{\cal F}|}\,[\Theta({\cal F})-{\rm i}\Theta(-{\cal F})]$ and $m^2=m^2-{\rm i}0^+$; $\Theta(\cdot)$ is the Heaviside function.
Moreover, the integration contour in Eqs.~\eqref{eq:pi_T}-\eqref{eq:pi_sFsF} is implicitly assumed to lie slightly above the real positive $s$ axis.
Especially for the cases of either a purely magnetic or electric field to be considered in the remainder of this section, the four-vectors $(kF)^\mu:=k_\alpha F^{\alpha\mu}$ and $(k{}^\star\!F)^\mu:=k_\alpha {}^\star \!F^{\alpha\mu}$ can be expressed as $(kF)^\mu=-(0,\vec{k}\times\vec{B})$,  $(k{}^\star\!F)^\mu=-(\vec{k}\cdot\vec{B},k^0\vec{B})$ and  $(kF)^\mu=-(\vec{k}\cdot\vec{E},k^0\vec{E})$, $(k{}^\star\!F)^\mu=(0,\vec{k}\times\vec{E})$, respectively.
In turn, for a purely magnetic or electric field our normalization of the tensor structures in \Eqref{eq:PiF} clearly ensures that these are independent of the amplitude profile of the background field and may depend only on its orientation.
Also note that in the constant field limit ($\partial_\alpha F^{\mu\nu}=0$ for all values of the indices $\alpha$, $\mu$ and $\nu$) the expression in the square brackets in \Eqref{eq:PiF} becomes independent of $x$. In this case the integration over space-time can be performed right away, resulting in an overall momentum conserving Dirac delta function that ensures $k'^\mu=k^\mu$.

Especially due to the rather complicated dependence of Eqs.~\eqref{eq:pi_T}-\eqref{eq:pi_sFsF} on the background field amplitude, in general the Fourier integral in \Eqref{eq:PiF} cannot be evaluated analytically for inhomogeneous fields.
On the other hand, the parameter integrals in Eqs.~\eqref{eq:pi_T}-\eqref{eq:pi_sFsF} can be performed in closed form \cite{Karbstein:2015cpa}. 
However, noting that the structure of Eqs.~\eqref{eq:PiF}-\eqref{eq:pi_sFsF} is such that the Fourier integral can be carried out explicitly for magnetic and electric fields with the Lorentzian amplitude profile in \Eqref{eq:calE}, in the present work we pursue this direction. 
The Fourier integral in a inhomogeneous direction simply amounts to an elementary Gaussian integral. 

Focusing on the case of a purely magnetic field, inserting the field profile~\eqref{eq:calE} into Eqs.~\eqref{eq:PiF}-\eqref{eq:pi_sFsF} and carrying out the Fourier integral, we arrive at
\begin{multline}
 \Pi^{\mu\nu}(-k',k)
 = (2\pi)^{4-d}\delta^{(4-d)}(k'-k)\biggl(\,\prod_{i=1}^d w_i\biggr)\Bigl(\frac{\pi}{4}\Bigr)^{\frac{d}{2}}\\
 \times\Bigl[
 \bigl((k'k)g^{\mu\nu} - k'^\mu k^\nu \bigr)\pi_T^d 
 + \frac{(k'F)^\mu  (kF)^\nu}{2{\cal F}}\,\pi_{FF}^d
 + \frac{(k'{}^\star\!F)^\mu (k{}^\star\!F)^\nu}{2{\cal F}}\,\pi_{{}^\star\!F{}^\star\!F}^d 
 \Bigr]\,, \label{eq:PiFd}
\end{multline}
where $\delta^{(4-d)}(k'-k)$ denotes the Dirac delta function in $4-d$ dimensions which ensures energy/momentum conservation in the homogeneous directions of the background field.
The scalar functions in \Eqref{eq:PiFd} are given by
\begin{align}
    \pi_T^d&=-\frac{\alpha}{2\pi}\int_0^\infty\frac{{\rm d}s}{s}\Bigl(\frac{e{\cal E}_0}{m^2}\frac{1}{s}\Bigr)^{\frac{d}{2}}{\rm e}^{-\frac{m^2}{e{\cal E}_0}s-\frac{e{\cal E}_0}{m^2}\frac{f(k',k)}{s}}\biggl(\frac{1}{\sinh^2 s}-\frac{\coth s}{s}+\frac{2}{3}\biggr)\,,\label{eq:pi_Td} \\
    \pi_{FF}^d&=-\frac{\alpha}{2\pi}\int_0^\infty\frac{{\rm d}s}{s}\Bigl(\frac{e{\cal E}_0}{m^2}\frac{1}{s}\Bigr)^{\frac{d}{2}}{\rm e}^{-\frac{m^2}{e{\cal E}_0}s-\frac{e{\cal E}_0}{m^2}\frac{f(k',k)}{s}}\biggl(\frac{1-2s\coth s}{\sinh^2 s}+\frac{\coth s}{s}\biggr)\,, \\
    \pi_{{}^\star\!F{}^\star\!F}^d&=-\frac{\alpha}{2\pi}\int_0^\infty\frac{{\rm d}s}{s}\Bigl(\frac{e{\cal E}_0}{m^2}\frac{1}{s}\Bigr)^{\frac{d}{2}}{\rm e}^{-\frac{m^2}{e{\cal E}_0}s-\frac{e{\cal E}_0}{m^2}\frac{f(k',k)}{s}}\biggl(\frac{1}{\sinh^2 s}-\frac{\coth s}{s}+\frac{2}{3}\,s\coth s\biggr)\,, \label{eq:pi_sFsFd}
\end{align}
with the dimensionless parameter
\begin{equation}
  f(k',k) := \frac{1}{16}\sum_{i=1}^d w_i^2(k'_i-k_i)^2 \geq0	\label{eq:fkks}
\end{equation}
measuring the momentum difference between the in and outgoing probe photon legs.
Nonzero values of \Eqref{eq:fkks} imply a finite momentum transfer from the inhomogeneous background to the probe field.
From the above discussion it is clear that the analogous result for a purely electric field follows from Eqs.~\eqref{eq:PiFd}-\eqref{eq:pi_sFsFd} by the substitution ${\cal E}_0\to-{\rm i}{\cal E}_0$.
Accounting for the fact that $(2\pi)^{(4-d)}\delta^{(4-d)}(k'-k=0)=\int{\rm d}^{4-d}x=:V^{(4-d)}$ amounts to the $(4-d)$-dimensional space-time volume associated with the homogeneous directions of the field inhomogeneity, from \Eqref{eq:PiFd} it is obvious that, in accordance with expectations, the $4$-dimensional space-time region $V^{(4-d)}(\prod_{i=1}^d w_i)$ where the background field can impact the propagation of probe light gets reduced with an increasing localization of the inhomogeneity.

We emphasize that within the constraints inherent to its determination via (i), Eqs.~\eqref{eq:PiFd}-\eqref{eq:fkks} allow for the study of photon propagation effects at one loop in field Lorentzian field inhomogeneities~\eqref{eq:calE} of arbitrarily strong peak field amplitudes. 
However, we first note that in the perturbative weak field limit characterized by $e{\cal E}_0/m^2\ll1$ the expressions in Eqs.~\eqref{eq:pi_Td}-\eqref{eq:pi_sFsFd} are amenable to closed-form all-order Taylor expansions.
To this end, it is convenient to shift the integration variable of the parameter integration as $s\to (e{\cal E}_0/m^2)s$ and make use of the power series for $\coth x$ about $x=0$; note that $1/\sinh^2 x=-\partial_x\coth x$.
Upon exchanging the all-order summation and the parameter integration over $s$, the latter can be performed explicitly and we obtain
\begin{equation}
 \begin{Bmatrix}
  \pi_{T}^d\\
  \pi_{FF}^d	\\
  \pi_{{}^\star\!F{}^\star\!F}^d
 \end{Bmatrix}
 =-\frac{\alpha}{\pi}\sum_{n=1}^{\infty} 2\bigl(\sqrt{f(k',k)}\,\bigr)^{2n-\frac{d}{2}} K_{2n-\frac{d}{2}} \bigl(2\sqrt{f(k',k)}\,\bigr)\frac{2\mathcal{B}_{2n+2}}{(2n+1)!}\left(\frac{2e{\cal E}_0}{m^2}\right)^{2n}
 \begin{Bmatrix}
  1	\\
  2n	\\
  1-\frac{2n+1}{6}\frac{\mathcal{B}_{2n}}{\mathcal{B}_{2n+2}}
 \end{Bmatrix}
	\label{eq:weakp}
\end{equation}
for a magnetic field and $d\geq1$. Here, $K_\nu(\cdot)$ denotes the modified Bessel function of the second kind and $\mathcal{B}_{n}$ are Bernoulli numbers.
Clearly, the analogous expansion for an electric field differs only by a factor of $(-1)^n$ in each summand. 
The modified Bessel functions ensure that \Eqref{eq:weakp} is strongly peaked at $f(k',k)\to0$, i.e., receives its main contribution for vanishing momentum transfer from the background field.
As to be expected \cite{Gies:2013yxa,Gies:2014wsa}, because of $K_\nu(z)=\sqrt{\pi/(2z)}\,{\rm e}^{-z}\bigl(1+{\cal O}(1/z)\bigr)$ the components of the polarization tensor receive an exponential suppression for large momentum transfers $f(k',f)\gg1$; see formula 10.25.3 of \cite{NIST}. 
Though \Eqref{eq:weakp} was derived for $d \geq 1$, it also allows to recover the uniform constant field result corresponding to $d=0$: namely, by taking the limit of $w_i\to\infty$ for all inhomogeneous dimensions $1\leq i\leq d$ and using the expansion of the modified Bessel function for small arguments (Ref.~\cite{NIST}: 10.30.2). From this we infer that
\begin{equation}
    (2\pi)^{4-d}\delta^{(4-d)}(k'-k) \biggl(\,\prod_{i=1}^d w_i\biggr)\Bigl(\frac{\pi}{4}\Bigr)^{\frac{d}{2}}2\bigl(\sqrt{f(k',k)}\,\bigr)^{2n-\frac{d}{2}} K_{2n-\frac{d}{2}} \bigl(2\sqrt{f(k',k)}\,\bigr) \xrightarrow[w_i \rightarrow \infty]{}(2\pi)^{4}\delta^{(4)}(k'-k) \, ,
\end{equation}
which recovers the well-known results in a uniform constant background field.
Moreover, we emphasize that in the special case where $k'_i=k_i$ in the inhomogeneous directions $1\leq i\leq d$, and thus $f(k',k)=0$, \Eqref{eq:weakp} can be substantially simplified by noting that $\lim_{f\to0}2(\sqrt{f})^{2n-\frac{d}{2}} K_{2n-\frac{d}{2}} \bigl(2\sqrt{f}\bigr)=\Gamma(2n-\tfrac{d}{2})$, where $\Gamma(\cdot)$ is the gamma function; see formula 10.30.2 of Ref.~\cite{NIST}.
As a direct consequence of Furry's theorem \cite{Furry:1937zz} (charge conjugation invariance of QED) the expansion in \Eqref{eq:weakp} is in even powers of $e{\cal E}_0$.
We remark that alternatively one could arrive at \Eqref{eq:weakp} by first carrying out an all-order weak-field expansion of the coefficients~\eqref{eq:pi_T}-\eqref{eq:pi_sFsF} of the photon polarization tensor in a constant magnetic/electric field, then replacing the amplitude profile by \Eqref{eq:calE} and finally performing the Fourier integral in \Eqref{eq:PiF}.
In line with this, with respect to the dependence on the peak amplitude ${\cal E}_0$ the structure of \Eqref{eq:weakp} is not surprising.

On the other hand, in the special case of $f(k',k)=0$, which is for instance determining the amplitude of strict forward scattering for probe photons with $k'^\mu=k^\mu$, the parameter integrations in Eqs.~\eqref{eq:pi_Td}-\eqref{eq:pi_sFsFd} can be performed explicitly for $0\leq d\leq3$, thereby allowing for analytical insights for arbitrary values of the peak field amplitude ${\cal E}_0$ encoded in the dimensionless parameter $e{\cal E}_0/m^2=:1/(2h)$ $\leftrightarrow$ $h=m^2/(2e{\cal E}_0)$ \cite{Dittrich:1978fc}.
For the case of a magnetic field inhomogeneity and $0\leq d\leq3$ the explicit results of these calculations can be expressed in terms of the Hurwitz zeta function $\zeta(s,h)$ and derivatives thereof; $\zeta'(s,h)=\partial_s\zeta(s,h)$, $\zeta(s)=\zeta(s,1)$ is the Riemann zeta function and $\psi(h)=\Gamma'(h)/\Gamma(h)$ is the digamma function. This results in the following rather compact representations,
\begin{equation}
 \begin{split}
  \pi_{T}^{d=0} &= -\frac{\alpha}{\pi}\biggl\{4\zeta'(-1,h)-h\bigl[2\zeta'(0,h)-\ln h+h\bigr]-\frac{1}{3}\ln h-\frac{1}{6}\biggr\}\, ,\\
  \pi_{FF}^{d=0} &= -\frac{\alpha}{\pi}\biggl\{\frac{1}{3}-h\bigl[2\zeta'(0,h)+\ln h+2h\bigl(1-\psi(h)\bigr)-1\bigr]\biggr\}\, ,\\
  \pi_{{}^\star\!F{}^\star\!F}^{d=0} &= -\frac{\alpha}{\pi}\biggl\{4\zeta'(-1,h)-h\bigl[2\zeta'(0,h)-\ln h +h\bigr]-\frac{1}{6}\bigl[2\psi(h)+h^{-1}+1\bigr]\biggr\}\, ,\label{eq:forward0D}
 \end{split}
\end{equation}
\begin{equation}
 \begin{split}
  \pi_{T}^{d=1} &= -\frac{2\alpha}{3\pi^{\frac{1}{2}}}h^{-\frac{1}{2}}\biggl\{-10\zeta(-\tfrac{3}{2},h)+6h\zeta(-\tfrac{1}{2},h)-h^{\frac{1}{2}}(1-2h)	\biggr\}\, ,\\
  \pi_{FF}^{d=1} &= -\frac{2\alpha}{3\pi^{\frac{1}{2}}}h^{-\tfrac{1}{2}}\biggl\{-5\zeta(-\tfrac{3}{2},h)+12h\zeta(-\tfrac{1}{2},h)-3h^2\zeta(\tfrac{1}{2},h)-2h^{\frac{3}{2}}\biggr\}\, ,\\
  \pi_{{}^\star\!F{}^\star\!F}^{d=1} &= -\frac{2\alpha}{3\pi^{\frac{1}{2}}}h^{-\frac{1}{2}}\biggl\{-10\zeta(-\tfrac{3}{2},h)+6h\zeta(-\tfrac{1}{2},h)+\frac{1}{2}\zeta(\tfrac{1}{2},h)-h^{-\frac{1}{2}}\Bigl(\frac{1}{4}-2h^2\Bigr)\biggr\}\, ,\label{eq:forward1D}
 \end{split}
\end{equation}
\begin{equation}
 \begin{split}
  \pi_{T}^{d=2} &= -\frac{\alpha}{\pi}h^{-1}\biggl\{-3\zeta'(-2,h)+h\biggl[2\zeta'(-1,h)-\frac{1}{2}h\ln h+\frac{1}{12}(2h^2+4\ln h+1)\biggr] \biggr\}\, ,\\
  \pi_{FF}^{d=2} &= -\frac{\alpha}{\pi}h^{-1}\biggl\{-3\zeta'(-2,h)+h\Bigl[6\zeta'(-1,h)-\frac{1}{12}-h(2\zeta'(0,h)-\ln h+\frac{5}{6}h)\Bigr]\biggr\}\, ,\\
  \pi_{{}^\star\!F{}^\star\!F}^{d=2} &= -\frac{\alpha}{\pi}h^{-1}\biggl\{-3\zeta'(-2,h)+\frac{1}{3}\zeta'(0,h)+h\biggl[2\zeta'(-1,h)+\frac{5}{12}+\frac{1}{6}h^2\biggr]+\frac{1}{6}(1-3h^2)\ln h\biggr\}\, , \label{eq:forward2D}
 \end{split}
\end{equation}
\begin{equation}
 \begin{split}
  \pi_{T}^{d=3} &= -\frac{4\alpha}{15\pi^{\frac{1}{2}}}h^{-\frac{3}{2}}\biggl\{14\zeta(-\tfrac{5}{2},h)-10h\zeta(-\tfrac{3}{2},h)+h^{\frac{3}{2}}\Bigl(\frac{5}{3}-2h\Bigr) \biggr\}\, ,\\
  \pi_{FF}^{d=3} &= -\frac{4\alpha}{15\pi^{\frac{1}{2}}}h^{-\frac{3}{2}}\biggl\{21\zeta(-\tfrac{5}{2},h)-40h\zeta(-\tfrac{3}{2},h)+15h^2\zeta(-\tfrac{1}{2},h)+2h^{\frac{5}{2}}\biggr\}\, ,\\
  \pi_{{}^\star\!F{}^\star\!F}^{d=3} &= -\frac{4\alpha}{15\pi^{\frac{1}{2}}}h^{-\frac{3}{2}}\biggl\{14\zeta(-\tfrac{5}{2},h)-10h\zeta(-\tfrac{3}{2},h)-\frac{5}{2}\zeta(-\tfrac{1}{2},h)+h^{\frac{1}{2}}\Bigl(\frac{5}{4}-2h^2\Bigr)\biggr\}\, .\label{eq:forward3D}
 \end{split}
\end{equation}
We point out that the uniform constant field $(d=0)$ result given in \Eqref{eq:forward0D} was already derived in Ref.~\cite{Karbstein:2015cpa}.
In line with the above discussion, the analogous results for an electric field follow via $h\to{\rm i}h$.

While an expansion of Eqs.~\eqref{eq:forward1D}-\eqref{eq:forward3D} for large $h\to\infty$ recovers the perturbative weak field expansion in \Eqref{eq:weakp}, an expansion about $h=0$ provides access to their strong field behavior. With the help of the series representation of the Hurwitz zeta function for small $h$ we readily obtain
\begin{equation}
 \begin{split}
  \pi_{T}^{d=0} &=-\frac{\alpha}{\pi}\biggl\{ -\frac{1}{3}\ln h+4\zeta'(-1)-\frac{1}{6}-h\ln h+\bigl[2-\ln(2\pi)\bigr]h+2\sum_{n=2}^{\infty}(-1)^n\frac{n-2}{n(n-1)}\zeta(n-1)h^n \biggr\}\, ,\\
  \pi_{FF}^{d=0} &=-\frac{\alpha}{\pi}\biggl\{\frac{1}{3}+h\ln h+\bigl[\ln(2\pi)-1\bigr]h-2h^2-2\sum_{n=3}^{\infty}(-1)^n\frac{n-2}{n-1}\zeta(n-1)h^n\biggr\}\, ,\\
  \pi_{{}^\star\!F{}^\star\!F}^{d=0} &=-\frac{\alpha}{\pi}\biggl\{\frac{1}{6}\frac{1}{h}+4\zeta'(-1)+\frac{1}{6}\left[2\gamma-1\right]-h\ln h+\Bigl[2-\ln(2\pi)-\frac{\zeta(2)}{6}\Bigr]h+\bigl[1+\zeta(3)\bigr]h^2 \\
  &\hspace*{5.2cm}+\sum_{n=3}^{\infty}(-1)^n\Bigl[\frac{2(n-2)}{n(n-1)}\zeta(n-1)+\frac{\zeta(n+1)}{6})\Bigr]h^n\biggr\}\, , \label{eq:forwardstrong0D}
 \end{split}
\end{equation}
\begin{equation}
 \begin{split}
  \pi_{T}^{d=1} &= -\frac{\alpha}{\pi}\biggl\{-\frac{2\sqrt{\pi}}{3}-\frac{4\pi^{\frac{3}{2}}}{3}h+\sum_{n=0}^{\infty}\frac{(-1)^n(2n-5)}{n!}\rho(n-\tfrac{3}{2})h^{n-\frac{1}{2}}\biggr\}\, ,\\
  \pi_{FF}^{d=1} &= -\frac{\alpha}{\pi}\biggl\{\frac{4\sqrt{\pi}}{3}h-\sum_{n=0}^{\infty}\frac{(-1)^n(2n-5)(n-\frac{1}{2})}{n!}\rho(n-\tfrac{3}{2})h^{n-\frac{1}{2}}\biggr\}\, ,\\
  \pi_{{}^\star\!F{}^\star\!F}^{d=1} &= -\frac{\alpha}{\pi}\biggl\{\frac{\sqrt{\pi}}{6}\frac{1}{h}-\frac{4\pi^{\frac{1}{2}}}{3}h+\sum_{n=0}^{\infty}(-1)^n\frac{(2n-5)\rho(n-\tfrac{3}{2})+\frac{1}{3}\rho(n+\tfrac{1}{2})}{n!}h^{n-\frac{1}{2}}\biggr\}\, , \label{eq:forwardstrong1D}
 \end{split}
\end{equation}
\begin{equation}
 \begin{split}
  \pi_{T}^{d=2} &= -\frac{\alpha}{\pi}\biggl\{\frac{3\zeta(3)}{4\pi^2}\frac{1}{h}+\frac{1}{3}\ln h-4\zeta'(-1)-\frac{1}{6}+\frac{1}{2}h\ln h+\frac{2\ln(2\pi)-5}{4}h-\frac{1}{3}h^2
  -2\sum_{n=3}^{\infty}(-1)^n\frac{n-2}{n+1}\zeta(n-1)h^n\biggr\},\hspace*{-1mm}\\
  \pi_{FF}^{d=2} &= -\frac{\alpha}{\pi}\biggl\{\frac{3\zeta(3)}{4\pi^2}\frac{1}{h}-\frac{1}{3}+\frac{3}{4}\bigl[1-2\ln(2\pi)\bigr]h+2\sum_{n=2}^{\infty}\frac{(-1)^n(n-2)}{(n-1)(n+1)}\zeta(n-1)h^n\biggr\}\, ,\\
  \pi_{{}^\star\!F{}^\star\!F}^{d=2} &= -\frac{\alpha}{\pi}\biggl\{-\frac{1}{6}\frac{1}{h}\ln h+\Bigl[\frac{3\zeta(3)}{4\pi^2}-\frac{\ln(2\pi)}{6}\Bigr]\frac{1}{h}+\frac{1-2\gamma-24\zeta'(-1)}{6}+\frac{1}{2}h\ln h+\Bigl[\ln(2\pi)+\frac{\pi^2}{18}-\frac{5}{2}\Bigr]h\\
  &\hspace*{6.2cm} -\sum_{n=2}^{\infty}\frac{(-1)^n}{n+1}\biggl[\frac{2(n-2)}{n(n-1)}\zeta(n-1)+\frac{1}{3}\zeta(n+1)\biggr]h^n\biggr\}\, ,\label{eq:forwardstrong2D}
 \end{split}
\end{equation}
\begin{equation}
 \begin{split}
  \pi_{T}^{d=3} &= -\frac{\alpha}{\pi}\biggl\{\frac{4}{9\pi^{\frac{5}{2}}}+\frac{8}{15\pi^{\frac{5}{2}}}h+\sum_{n=0}^{\infty}\frac{(-1)^n(2n-7)}{n!}\rho(n-\tfrac{5}{2})h^{n-\frac{3}{2}}\biggr\}\, ,\\
  \pi_{FF}^{d=3} &= -\frac{\alpha}{\pi}\biggl\{-\frac{8}{15\pi^{\frac{5}{2}}}h-\sum_{n=0}^{\infty}\frac{(-1)^n(2n-7)(n-\frac{3}{2})}{n!}\rho(n-\tfrac{5}{2})h^{n-\frac{3}{2}}\biggr\}\, ,\\
  \pi_{{}^\star\!F{}^\star\!F}^{d=3} &= -\frac{\alpha}{\pi}\biggl\{-\frac{\sqrt{\pi}}{3}\frac{1}{h}+\frac{8}{15\pi^{\frac{5}{2}}}h+\sum_{n=0}^{\infty}(-1)^n\frac{(n-\frac{7}{2})\rho(n-\tfrac{5}{2})+\frac{1}{6}\rho(n-\tfrac{1}{2})}{n!}h^{n-\frac{3}{2}}\biggr\}\, , \label{eq:forwardstrong3D}
 \end{split}
\end{equation}
where $\gamma$ is the Euler–Mascheroni constant and we used the shorthand notation $\rho(s) := \zeta(s)\Gamma(s) = \int_0^\infty{\rm d}t\,t^{s-1}/(\mathrm{e}^t-1)$.
We emphasize that the contributions in Eqs.~\eqref{eq:forwardstrong0D}-\eqref{eq:forwardstrong3D} are not necessarily ordered with respect to their importance in the strong field limit. Instead, aiming at providing most compact representations, we have accounted for as many terms in the infinite sums as possible.
For completeness, we also note that the leading contribution to $\pi_T^{d=0}$ in \Eqref{eq:forwardstrong0D} is of the form $\lim_{h\to0}\pi_T^{d=0}=-\alpha\beta_1\ln(e{\cal E}_0/m^2)$, where $\beta_1=1/(3\pi)$ denotes the leading coefficient of the QED $\beta$ function governing the running of the fine structure constant.

From Eqs.~\eqref{eq:forwardstrong0D}-\eqref{eq:forwardstrong3D} we infer that with an increasing localization of the inhomogeneity the leading strong field behavior of a given component $\pi_p^d$ with $p\in\{T,FF,{}^\star\!F{}^\star\!F\}$ becomes more pronounced: counting $\ln h\sim{\cal O}(h^0)$, an increase of the number of inhomogeneous directions from $d$ to $d+1$ generically comes with an enhancement by a factor $\sim h^{-1/2}=(2e{\cal E}_0/m^2)^{1/2}$.
For $0\leq d\leq 2$ the scaling of $\pi_T^d$ and $\pi_{FF}^d$ with $h\ll1$ is subleading in comparison to $\pi_{{}^\star\!F{}^\star\!F}^d$. In the case of $d=2$ the scaling of the latter is just logarithmically enhanced by a factor of $\ln(1/h)\sim\ln(e{\cal E}_0/m^2)$.
Interestingly, for $d=3$ all components in \Eqref{eq:forwardstrong3D} exhibit the same leading scaling $\sim h^{-3/2}$.

For insights into $e{\cal E}_0/m^2\gg1$ and $k'^\mu\neq k^\mu$ it is convenient to invoke the same substitution as in the determination of the perturbative weak field limit~\eqref{eq:weakp}, but then instead to perform an expansion of the expressions in the round brackets in Eqs.~\eqref{eq:pi_Td} -\eqref{eq:pi_sFsFd} containing the hyperbolic functions for $se{\cal E}_0/m^2\gg1$.
This yields
\begin{equation}
 \begin{Bmatrix}
  \pi_{T}^d\\
  \pi_{FF}^d	\\
  \pi_{{}^\star\!F{}^\star\!F}^d
 \end{Bmatrix}
 =-\frac{\alpha}{\pi}\int_0^\infty\frac{{\rm d}s}{s}\Bigl(\frac{1}{s}\Bigr)^{\frac{d}{2}}{\rm e}^{-s-\frac{f(k',k)}{s}}\left[
 \begin{Bmatrix}
  \frac{1}{3}-\frac{h}{s}	\\
  \frac{h}{s}	\\
  \frac{1}{6}\frac{s}{h}-\frac{h}{s}
 \end{Bmatrix}
 +{\cal O}\bigl({\rm e}^{-s/h}\bigr)\right].
	\label{eq:eEbym2timesflarge}
\end{equation}
which, given that $f(k',k)$ in \Eqref{eq:fkks} is large enough such as to sufficiently dampen the potentially divergent contributions multiplying the exponential factors in the integrand for $s\to0$, allows for a reliable approximation. 
The latter assumption is increasingly well justified for large values of $f(k',k)$, while the limit of $f(k',k) \to 0$ has to be handled with care.
Upon performing the integration in \Eqref{eq:eEbym2timesflarge} and keeping the leading terms for $h\ll1$ only, we obtain
\begin{equation}
 \begin{Bmatrix}
  \pi_{T}^d\\
  \pi_{FF}^d	\\
  \pi_{{}^\star\!F{}^\star\!F}^d
 \end{Bmatrix}
 \simeq-\frac{\alpha}{3\pi}\frac{1}{h}
 \begin{Bmatrix}
  2h\bigl(\sqrt{f(k',k)}\,\bigr)^{-\frac{d}{2}} K_\frac{d}{2}\bigl(2\sqrt{f(k',k)}\bigr)	\\
   6h^2\bigl(\sqrt{f(k',k)}\,\bigr)^{-(\frac{d}{2}+1)} K_{\frac{d}{2}+1}\bigl(2\sqrt{f(k',k)}\bigr)\\
  \bigl(\sqrt{f(k',k)}\,\bigr)^{-(\frac{d}{2}-1)} K_{\frac{d}{2}-1}\bigl(2\sqrt{f(k',k)}\bigr) 
 \end{Bmatrix}.
	\label{eq:eEbym2timesflargeLO}
\end{equation}
Equation~\eqref{eq:eEbym2timesflargeLO} in particular implies that for sufficiently large momentum transfers from the field inhomogeneity~\eqref{eq:calE} the leading strong-field scaling of the components of the polarization tensor with $h\ll1$ is independent of the number of inhomogeneous directions $1\leq d\leq3$. Because in uniform constant fields the polarization tensor is nonzero only for $k'^\mu=k^\mu$ considering this limit for $d=0$ makes no sense.

In slowly varying magnetic-like background fields fulfilling ${\cal F}\geq0$ and for low-frequency photons the photon polarization tensor is real valued, signalizing the impossibility of electron-positron pair production under these conditions.
On the other hand, in electric-like fields for which ${\cal F}<0$ the polarization tensor features an imaginary part and pair production becomes possible.
For its determination we specialize Eqs.~\eqref{eq:pi_Td}-\eqref{eq:pi_sFsFd} to an electric field inhomogeneity via the replacement ${\cal E}_0\to-{\rm i}{\cal E}_0$. Substituting $s\to-{\rm i}\tilde s$ and deforming the integration contour such as to lie slightly below the real positive $\tilde s$ axis, the associated imaginary parts can then be worked out by noting that ${\rm Im}\{\pi_p^d\}=[\pi_p^d-(\pi_p^d)^*]/2{\rm i}$.
The latter identity maps the determination of the imaginary part of $\pi_p^d$ to performing a contour integral in the complex $\tilde s$ plane enclosing the real positive $\tilde s$ axis, which can be readily evaluated with the Cauchy's residue theorem.
This yields the exact expression
\begin{multline}
    {\rm Im}\begin{Bmatrix}
    \pi_T^d \\
    \pi_{FF}^d	\\
    \pi_{{}^\star\!F{}^\star\!F}^d
    \end{Bmatrix}
    = \frac{\alpha}{2\pi^2}\sum_{n=1}^{\infty}\Bigl(\frac{e{\cal E}_0}{m^2}\frac{1}{n\pi}\Bigr)^{\frac{d}{2}}\,\frac{\mathrm{e}^{-\frac{m^2}{e{\cal E}_0}n\pi-\frac{e{\cal E}_0}{m^2}\frac{f(k',k)}{n\pi}}}{n^2}
    \\ \times
    \begin{Bmatrix}
        2+\frac{d}{2}+\frac{m^2}{e{\cal E}_0}n\pi-\frac{e{\cal E}_0}{m^2}\frac{f(k',k)}{n\pi} \\
        \frac{d}{2}\left(2+\frac{d}{2}\right)+(1+d)\bigl(\frac{m^2}{e{\cal E}_0}n\pi-\frac{e{\cal E}_0}{m^2}\frac{f(k',k)}{n\pi}\bigr)+\bigl(\frac{m^2}{e{\cal E}_0}n\pi-\frac{e{\cal E}_0}{m^2}\frac{f(k',k)}{n\pi}\bigr)^2-\frac{2}{n\pi}\frac{e{\cal E}_0}{m^2}\frac{f(k',k)}{n\pi} \\
        2+\frac{d}{2}+\frac{2}{3}(n\pi)^2+\frac{m^2}{e{\cal E}_0}n\pi-\frac{e{\cal E}_0}{m^2}\frac{f(k',k)}{n\pi}
    \end{Bmatrix}\, .\label{IfulldD}
\end{multline}
Conversely, for the case of a magnetic field the hyperbolic functions in Eqs.~\eqref{eq:pi_Td}-\eqref{eq:pi_sFsFd} do not feature any poles on the real positive $s$ axis making the analogous contour integral vanish. In turn no imaginary part appears in that case.
We emphasize that \Eqref{IfulldD} is a manifestly non-perturbative result: it is characterized by the same non-perturbative exponent as the Schwinger effect \cite{Schwinger:1951nm}.
Correspondingly, it cannot be obtained by performing a naive perturbative expansion, such as adopted in \Eqref{eq:weakp}.
In the weak electric field limit $e{\cal E}_0/m^2\ll1$ the leading contributions to \Eqref{IfulldD} arise from the $n=1$ term.
Focusing on $k'^\mu=k^\mu$, which is the limit relevant for the determination of the single photon assisted pair production rate in the electric background field (cf. Sec.~\ref{sec:phys} below), and keeping only the leading terms we then find
\begin{equation}
    {\rm Im}\begin{Bmatrix}
    \pi_T^d \\
    \pi_{FF}^d	\\
    \pi_{{}^\star\!F{}^\star\!F}^d
    \end{Bmatrix}
    \simeq \frac{\alpha}{2\pi^2}\Bigl(\frac{e{\cal E}_0}{m^2}\frac{1}{\pi}\Bigr)^{\frac{d}{2}-1}\,\mathrm{e}^{-\frac{m^2}{e{\cal E}_0}\pi}\,
    \begin{Bmatrix}
        1 \\
        \frac{m^2}{e{\cal E}_0}\pi \\
        1
    \end{Bmatrix}\, .\label{eq:ImPipweakfield}
\end{equation}
Equation~\eqref{eq:ImPipweakfield} implies that in weak electric fields the imaginary part of the polarization tensor decreases with increasing localization of the background field: similarly as for the Schwinger effect in the absence of additional photons \cite{Karbstein:2017pbf}, each additional inhomogeneous direction in \Eqref{eq:calE} comes with a reduction by a factor of $(e{\cal E}_0/m^2)^{1/2}\ll1$.

Together with \Eqref{eq:weakp} specialized to the electric field case forming its real part, \Eqref{eq:ImPipweakfield} constitutes the full result of the photon polarization tensor in the electric field inhomogeneity~\eqref{eq:calE}.
On the other hand, the strong field expansions performed in Eqs.~\eqref{eq:forwardstrong0D}-\eqref{eq:forwardstrong3D} and Eqs. \eqref{eq:eEbym2timesflarge}-\eqref{eq:eEbym2timesflargeLO} provide direct access to both the real and imaginary parts of the polarization tensor in strong electric fields $e{\cal E}_0/m^2\gg1$.
In fact, alternatively the strong field $e{\cal E}_0/m^2\gg1$ limits for the imaginary part for both $k^\mu=k'^\mu$ and sufficiently large values of $f(k,k')$ (cf. also above) can be extracted directly from \Eqref{IfulldD}.

For $f(k',k) = 0$, the sums can be performed explicitly yielding polylogarithms (Ref.~\cite{NIST}: 25.12.10), the expansions of which for $e{\cal E}_0/m^2\gg1$ follow from formula 25.12.12 of \cite{NIST}. This allows us to readily infer that the leading terms are given by
\begin{equation}
    {\rm Im}\begin{Bmatrix}
  \pi_{T}^d\\
  \pi_{FF}^d	\\
  \pi_{{}^\star\!F{}^\star\!F}^d
 \end{Bmatrix}
    = \frac{\alpha}{2\pi^2}\Bigl(\frac{e{\cal E}_0}{\pi m^2}\Bigr)^{\frac{d}{2}}
    \begin{Bmatrix}
        \bigl(2 + \frac{d}{2}\bigr)\zeta(2+\tfrac{d}{2}) \\
        \frac{d}{2}\bigl(2+\frac{d}{2}\bigr)\zeta(2+\tfrac{d}{2})+\delta_{0,d}\,\pi\frac{m^2}{e{\cal E}_0}\ln\frac{e{\cal E}_0}{m^2}\\
        \frac{2}{3}\pi^2h_d(\frac{e{\cal E}_0}{m^2})+\bigl(2+\frac{d}{2}\bigr)\zeta(2+\tfrac{d}{2})
    \end{Bmatrix}\, ,
    \label{eq:ImSF_LO}
\end{equation}
where $\delta_{i,j}$ denotes the Kronecker and we introduced the $d$ dependent quantities $h_0(\frac{e{\cal E}_0}{m^2}) = \frac{1}{\pi}\frac{e{\cal E}_0}{m^2}$, $h_1(\frac{e{\cal E}_0}{m^2}) = (\frac{e{\cal E}_0}{ m^2})^{1/2}$, $h_2(\frac{e{\cal E}_0}{m^2}) = \ln\frac{e{\cal E}_0}{m^2}$ and $h_3(\frac{e{\cal E}_0}{m^2}) = \zeta(\frac{3}{2})$. Note that with increasing values of $d$ the strong-field scaling of $h_d(\frac{e{\cal E}_0}{m^2})$ becomes less pronounced. While it ensures the component $\pi_{{}^\star\!F{}^\star\!F}^d$ to exhibit the leading strong-field behavior for $0\leq d\leq2$, this component becomes as important as the other ones for $d=3$.
As to be expected, \Eqref{eq:ImSF_LO} recovers the leading contributions to the imaginary part of the expressions in Eqs.~\eqref{eq:forwardstrong0D}-\eqref{eq:forwardstrong3D} specialized to an electric field inhomogeneity.

On the other hand, in the case of sufficiently large $f(k',k)$, where the summands are manifestly finite for $n\to0$ by assumption, we can make use of the fact that the $n$ dependence of \Eqref{IfulldD} is effectively in terms of the combined parameter $\frac{m^2}{e{\cal E}_0}n\pi$ and approximate the infinite sum by an integral via $\sum_{n=1}^\infty g(\frac{m^2}{e{\cal E}_0}n\pi)\to \frac{e{\cal E}_0}{m^2}\frac{1}{\pi}\int_0^\infty{\rm d}\nu\,g(\nu)$.
Adopting this strategy to \Eqref{IfulldD} and limiting ourselves only to the leading terms, we arrive at
\begin{equation}
 {\rm Im}\begin{Bmatrix}
  \pi_{T}^d\\
  \pi_{FF}^d	\\
  \pi_{{}^\star\!F{}^\star\!F}^d
 \end{Bmatrix}
 \simeq\frac{\alpha}{3\pi}\frac{e{\cal E}_0}{m^2}
 \begin{Bmatrix}
  6(\frac{m^2}{e{\cal E}_0})^{2}\bigl(\sqrt{f(k',k)}\,\bigr)^{-(\frac{d}{2}+1)} K_{\frac{d}{2}+1}\bigl(2\sqrt{f(k',k)}\bigr)	\\
  -3(\frac{m^2}{e{\cal E}_0})^{2}\bigl(\sqrt{f(k',k)}\,\bigr)^{-(\frac{d}{2}+1)} K_{\frac{d}{2}+1}\bigl(2\sqrt{f(k',k)}\bigr)\\
  2\bigl(\sqrt{f(k',k)}\,\bigr)^{-(\frac{d}{2}-1)} K_{\frac{d}{2}-1}\bigl(2\sqrt{f(k',k)}\bigr)
 \end{Bmatrix},
\end{equation}
which indeed recovers the imaginary part of Eqs.~\eqref{eq:eEbym2timesflarge}-\eqref{eq:eEbym2timesflargeLO}. Note that in order to obtain the result for $\pi_T^d$ given here from \Eqref{eq:eEbym2timesflarge} one has to account for the next to leading order contribution; the leading term given in \Eqref{eq:eEbym2timesflargeLO} is purely real valued.

\subsection{Crossed field}\label{subsec:crossed}

The exact expression for the one-loop photon polarization tensor in a constant crossed field fulfilling ${\cal F}={\cal G}=0$ which accounts for arbitrary momentum transfers $k^\mu$ through the charged particle loop \cite{Narozhny:1968} can be expressed as
\begin{equation}
 \Pi^{\mu\nu}(-k',k)
 =(2\pi)^4\delta^{(4)}(k'-k) \biggl[
 \bigl(k^2g^{\mu\nu} - k^\mu k^\nu \bigr)\pi_T 
 + \frac{(kF)^\mu  (kF)^\nu}{(kF)^2}\,\tilde\pi_{FF}
 + \frac{(k{}^\star\!F)^\mu (k{}^\star\!F)^\nu}{(kF)^2}\,\tilde\pi_{{}^\star\!F{}^\star\!F} 
 \biggr], \label{eq:PiCrossed}
\end{equation}
with scalar functions
\begin{equation}
    \begin{Bmatrix}
        \pi_{T} \\
        \tilde\pi_{FF} \\
        \tilde\pi_{{}^\star\!F{}^\star\!F}
    \end{Bmatrix}
    =\frac{\alpha}{2\pi}\int_0^\infty \frac{\mathrm{d}s}{s} \int_0^1 \mathrm{d}\nu \, \mathrm{e}^{-\mathrm{i}\bigl[\phi_0 + \frac{1}{4}(1-\nu^2)^2\frac{e^2(kF)^2s^2}{12m^6}\bigr]s}
    \begin{Bmatrix}
        -\mathrm{i}s(3-\nu^2)\nu^2\bigl[\frac{2}{3}\frac{k^2}{4m^2}+\frac{1}{3}(1-\nu^2)\frac{e^2(kF)^2s^2}{12m^6}\bigr] \\
        m^2(3+\nu^2)(1-\nu^2)\frac{e^2(kF)^2s^2}{12m^6} \\
        2m^2(3-\nu^2)(1-\nu^2)\frac{e^2(kF)^2s^2}{12m^6}
    \end{Bmatrix},
    \label{eq:pi-comps}
\end{equation}
where $s$ is the propertime, $\nu$ governs the momentum distribution in the charged particle loop, and
\begin{equation}
    \phi_0 = 1-{\rm i}0^+ + (1-\nu^2)\frac{k^2}{4m^2}
    \label{eq:PiCrossedPhase}
\end{equation}
is a dimensionless phase.
Equations~\eqref{eq:PiCrossed}-\eqref{eq:PiCrossedPhase} follow from the expressions given in equations (2.114) and (2.115) of \cite{Dittrich:2000zu} with the help of the identity (A.1) of \cite{Karbstein:2013ufa}.
Note that the entire four-momentum dependence of \Eqref{eq:pi-comps} is in terms of $k^2$ and $(kF)^2$. Upon introducing $\hat\kappa^\mu=(1,\hat{\vec{\kappa}})$ with $\hat{\vec{\kappa}}=(\vec{E}\times\vec{B})/(|\vec{B}||\vec{E}|)$, in constant crossed fields of amplitude ${\cal E}=|\vec{B}|=|\vec{E}|$ the latter expression can be compactly represented as $(kF)^2={\cal E}^2(\hat\kappa k)^2\geq0$.
The overall energy and momentum conserving delta function in \Eqref{eq:PiCrossed} reflects the fact that a constant electromagnetic field cannot transfer energy and momentum to the charged particle loop.
We emphasize that the structure of \Eqref{eq:PiCrossed} is very similar to \Eqref{eq:PiF} in the constant field limit. However, because of ${\cal F}=0$ in crossed fields, here we employ a slightly different normalization to render the tensor structures $\sim(kF)^\mu(kF)^\nu$ and $\sim(k{}^\star\!F)^\mu (k{}^\star\!F)^\nu$ independent of the field strength.
This immediately implies that in the present section their coefficients have a slightly different meaning;
to make this evident we decorate them with a tilde.
In particular note that while $\pi_T$ is dimensionless we have $\tilde\pi_{FF}\sim\tilde\pi_{{}^\star\!F{}^\star\!F}\sim m^2$.
For later reference we also note that making use of the substitution $se\sqrt{(kF)^2/m^6}\,(1-\nu^2)/(4\sqrt{3})\to s$, \Eqref{eq:pi-comps} can alternatively be expressed as
\begin{equation}
    \begin{Bmatrix}
        \pi_{T} \\
        \tilde\pi_{FF} \\
        \tilde\pi_{{}^\star\!F{}^\star\!F}
    \end{Bmatrix}
    =\frac{\alpha}{\pi}\int_0^\infty \mathrm{d}s \int_0^1 \mathrm{d}\nu \, \mathrm{e}^{-\mathrm{i}(\phi_0 + s^2)\frac{4\sqrt{3}}{1-\nu^2}\frac{m^3}{e\sqrt{(kF)^2}}s}
    \begin{Bmatrix}
        -\mathrm{i}\frac{4\sqrt{3}}{1-\nu^2}\frac{m^3}{e\sqrt{(kF)^2}}(3-\nu^2)\nu^2\bigl(\frac{1}{3}\frac{k^2}{4m^2}+\frac{1}{3}\frac{2}{1-\nu^2}s^2\bigr) \\
        2m^2\frac{3+\nu^2}{1-\nu^2}s \\
        4m^2\frac{3-\nu^2}{1-\nu^2}s
    \end{Bmatrix}.
    \label{eq:pi-comps2}
\end{equation}

Because of its specific tensor structure which is a direct consequence of $k'^\mu=k^\mu$ in constant fields, unfortunately \Eqref{eq:PiCrossed} cannot be adopted to the study of most generic situations where $k'^\mu\neq k^\mu$.
However, it should allow for reliable insights in the specific limit of $k'^\mu=k^\mu$ also for slowly varying inhomogeneous background fields.
To this end, we first identically rewrite the delta function in \Eqref{eq:PiCrossed} as $(2\pi)^4\delta^{(4)}(k'-k)=(2\pi)^{4-d}\delta^{(4-d)}(k'-k)\int{\rm d}^dx\,{\rm e}^{{\rm i}\sum_{i=1}^d(k_i-k'_i)x_i}$ and subsequently substitute the uniform constant field profile for a slowly varying inhomogeneous one with $d$ inhomogeneous directions.
Upon setting $k'^\mu=k^\mu$ for the momentum components in the $d$ inhomogeneous directions, determining the explicit expression for the photon polarization tensor in the inhomogeneous field in this particular limit then only requires performing an integration over coordinate space.
Specializing the study to a crossed field with amplitude profile~\eqref{eq:calE}, similar as in Sec.~\ref{subsec:B+E} this integration can be readily performed analytically for all scalar functions $\pi_T$, $\tilde\pi_{FF}$ and $\tilde\pi_{{}^\star\!F{}^\star\!F}$ determining the polarization tensor.
This becomes particularly obvious from \Eqref{eq:pi-comps2} where the field dependence in the exponential $\sim1/\sqrt{(kF)^2}\sim1/{\cal E}(x)$ and in the factor multiplying this exponential for $\pi_T$ ensures that only elementary Gaussian integrals are to be performed for Lorentzian amplitude profiles~\eqref{eq:calE}.

Carrying out the space-time integrations over the $d$ inhomogeneous directions, we obtain 
\begin{multline}
 (2\pi)^{(d)}\delta(k'-k)\,\Pi^{\mu\nu}(-k',k)
 = (2\pi)^{4}\delta^{(4)}(k'-k)\biggl(\,\prod_{i=1}^d w_i\biggr)\Bigl(\frac{\pi}{4}\Bigr)^{\frac{d}{2}}\\
 \times\Bigl[
 \bigl(k^2g^{\mu\nu} - k^\mu k^\nu \bigr)\pi_T^d 
 + \frac{(kF)^\mu  (kF)^\nu}{(kF)^2}\,\tilde\pi_{FF}^d
 + \frac{(k{}^\star\!F)^\mu (k{}^\star\!F)^\nu}{(kF)^2}\,\tilde\pi_{{}^\star\!F{}^\star\!F}^d 
 \Bigr]\,, \label{eq:PiFcrossd}
\end{multline}
where we multiplied both sides with the delta function $(2\pi)^{(d)}\delta(k'-k)$ to signalize and ensure that only the contribution for which $k'^\mu=k^\mu$ in the $d$ inhomogeneous directions is to be considered here.
The explicit expressions for the scalar functions in \Eqref{eq:PiFcrossd} are
\begin{equation}
    \begin{Bmatrix}
        \pi_{T}^d \\
        \tilde\pi_{FF}^d \\
        \tilde\pi^d_{{}^\star\!F{}^\star\!F}
    \end{Bmatrix}
    =\frac{\alpha}{\pi}\int_0^\infty \mathrm{d}s \int_0^1 \mathrm{d}\nu \, \frac{\mathrm{e}^{-\mathrm{i}\Phi_0s}}{(\mathrm{i}\Phi_0s)^{\frac{d}{2}}}
    \begin{Bmatrix}
        -\mathrm{i}\frac{4\sqrt{3}}{1-\nu^2}\frac{1}{\chi_0}\bigl(1+\frac{d}{2}\frac{1}{\mathrm{i}\Phi_0s}\bigr)(3-\nu^2)\nu^2\bigl(\frac{1}{3}\frac{k^2}{4m^2}+\frac{1}{3}\frac{2}{1-\nu^2}s^2\bigr) \\
        2m^2\frac{3+\nu^2}{1-\nu^2}s \\
        4m^2\frac{3-\nu^2}{1-\nu^2}s
    \end{Bmatrix},
    \label{eq:piFcrossd}
\end{equation}
where we introduced the {\it quantum nonlinear parameter} (cf., e.g., Ref.~\cite{Fedotov:2022ely}) of the peak field\footnote{Here, $F_0^{\mu\nu}$ denotes the field strength tensor of a constant crossed field of amplitude ${\cal E}_0$.}
\begin{equation}
    \chi_0=\sqrt{\frac{e^2(kF_0)^2}{m^6}}=\frac{e{\cal E}_0|\hat\kappa k|}{m^3}\,, \label{eq:chi0}
\end{equation}
and made use of the shorthand notation
\begin{equation}
    \Phi_0 = (\phi_0 + s^2)\frac{4\sqrt{3}}{1-\nu^2}\frac{1}{\chi_0}\,.
\end{equation}
In the next step, we note that the identity $({\rm i}\Phi_0s)^{-d/2}=1/\Gamma(d/2)\int_0^\infty{\rm d}t\,t^{d/2-1}\,{\rm e}^{-({\rm i}\Phi_0s)t}$ can be employed to recast \Eqref{eq:piFcrossd} in a form where its entire dependence on ${\rm i}\Phi_0s$ appears only in the argument of the exponential function.
Recalling the definition of Scorer's function \cite{Scorer:1950}, $\mathrm{Hi}(z) = \frac{1}{\pi}\int_0^\infty\mathrm{d}t\, \mathrm{e}^{-\frac{1}{3}t^3+zt}$, and rewriting it as
\begin{equation}
    {\rm Hi}\bigl(-{\rm i}z(3{\rm i}a)^{-\frac{1}{3}}\bigr) = (3{\rm i}a)^{\frac{1}{3}}\frac{1}{\pi}\int_0^\infty\mathrm{d}s\, \mathrm{e}^{-{\rm i}(z+as^2)s}\quad \text{for}\quad a>0\,, \label{eq:Hi}
\end{equation}
it is then obvious that the propertime integration in \Eqref{eq:piFcrossd} can be carried out explicitly and represented in terms of ${\rm Hi}(z)$ and derivatives thereof; we use the notation ${\rm Hi}'(z)={\rm d}{\rm Hi}(z)/{\rm d}z$.
In turn, for $1\leq d\leq3$ \Eqref{eq:piFcrossd} can be expressed as 
\begin{equation}
  \pi_T^d = -\frac{\alpha}{6} \frac{1}{\Gamma(\frac{d}{2})}\int_0^1 \mathrm{d}\nu\, \nu^2(3-\nu^2)\Bigl(\frac{4}{1-\nu^2}\Bigr)^{\frac{2}{3}}\int_0^\infty\frac{{\rm d}t}{t}\, \frac{t^{\frac{d}{2}}}{(1+t)^{\frac{2}{3}}}\Bigl(1+\frac{d}{2}t\Bigr) \biggl[2\mathrm{e}^{\mathrm{i}\frac{\pi}{3}}\frac{k^2}{4m^2}\,\chi_0^{-\frac{2}{3}}\,\mathrm{Hi}(\rho)+\frac{1}{3}\Bigl(\frac{4}{1-\nu^2}\Bigr)^{\frac{1}{3}}\,\mathrm{Hi}''(\rho)\biggr]
  \label{eq:crossedTHi}
 \end{equation}
 and
\begin{equation}
    \begin{Bmatrix}
     \tilde\pi_{FF}^d \\
     \tilde\pi^d_{{}^\star\!F{}^\star\!F}
 \end{Bmatrix}
 = \frac{\alpha}{6}\frac{\mathrm{e}^{-\mathrm{i}\frac{\pi}{3}}}{\Gamma(\frac{d}{2})} m^2 \chi_0^{\frac{2}{3}} \int_0^1 \mathrm{d}\nu\, \Bigl(\frac{4}{1-\nu^2}\Bigr)^{\frac{1}{3}}\int_0^\infty\frac{{\rm d}t}{t}\, \frac{t^{\frac{d}{2}}}{(1+t)^{\frac{2}{3}}}
 \begin{Bmatrix}
     3+\nu^2 \\
     2(3-\nu^2)
 \end{Bmatrix} \mathrm{Hi}'(\rho) \, ,
 \label{eq:crossedFHi}
\end{equation}
where the argument of Scorer's function and its derivatives depends on both parameters $\nu$ and $t$ to be integrated over, and is given by
\begin{equation}
\rho=\mathrm{e}^{- \mathrm{i}\frac{2\pi}{3}}\Bigl(\frac{1+t}{\chi_0}\frac{4}{1-\nu^2}\Bigr)^{\frac{2}{3}}\phi_0\,. \label{eq:rho}
\end{equation}
At this point, we also emphasize that the uniform constant field ($d=0$) results can be represented in a form very similar to those for $d>0$ given in Eqs.~\eqref{eq:crossedTHi}-\eqref{eq:rho}. They can be recovered from these expressions by omitting the integral over $t$, $\int_0^\infty{\rm d}t\to 1$, then setting $t^{d/2-1}/\Gamma(d/2)\to1$, and finally identifying $t\to0$ in all the other terms. 
Aside from two multiplicative factors of $k^2/(4m^2)\,\chi_0^{-2/3}$ and $\chi_0^{2/3}$, respectively, the entire dependence of Eqs.~\eqref{eq:crossedTHi} and \eqref{eq:crossedFHi} on the field amplitude and the momentum is encoded in the argument $\rho$ of Scorer's function.
This considerably simplifies the determination of the 
weak and strong field behavior discussed below.

Clearly, the weak field limit amounts to $\chi_0\ll1$ and thus follows from an expansion of Eqs.~\eqref{eq:crossedTHi} and \eqref{eq:crossedFHi} for $\rho\to\infty$. In turn, the corresponding result can be readily obtained from the power series of Scorer's function ${\rm Hi}(\rho)$ for large arguments (Ref.~\cite{NIST}: 9.12.27); the power series for ${\rm Hi}'(\rho)$ and ${\rm Hi}''(\rho)$ follow by differentiation.
Given that $4|1+k^2/(4m^2)|/\chi_0^{2/3
}\gg1$ holds,
the condition $|\rho|\gg1$ is fulfilled for all relevant values of the integration parameters $0\leq\nu\leq1$ and $0\leq t\leq\infty$.
Therewith, we immediately obtain the following all-order perturbative weak field expansions, 
\begin{multline}
    \pi_{T}^d = 
    -\frac{2\alpha}{\pi}\sum_{n=1}^{\infty}\frac{(3n-1)!\bigl[\Gamma(2n-\frac{d}{2})+\frac{d}{2}\Gamma(2n-1-\frac{d}{2})\bigr]}{48^n \Gamma(2n)\Gamma(n)} \\
    \times\int_{0}^{1}\mathrm{d}\nu\,\frac{\nu^2(3-\nu^2)(1-\nu^2)^{2n-1}}{\phi_0^{3n-2}}\Bigl[ \frac{1}{3} +\frac{k^2}{4m^2}\frac{1-\nu^2}{\phi_0^3}\frac{1}{2(3n-2)(3n-1)} \Bigr] \chi_0^{2n}
    \label{eq:weakT}
\end{multline} 
and
\begin{equation}
 \begin{Bmatrix}
     \tilde\pi_{FF}^d \\
     \tilde\pi^d_{{}^\star\!F{}^\star\!F}
 \end{Bmatrix}
 =-\frac{2\alpha}{\pi}m^2\sum_{n=1}^{\infty} \frac{(3n-2)!\,\Gamma(2n-\frac{d}{2})}{48^{n} \Gamma(2n)\Gamma(n)}\int_{0}^{1}\mathrm{d}\nu\,\frac{(1-\nu^2)^{2n-1}}{\phi_0^{3n-1}}
 \begin{Bmatrix}
  3+\nu	\\
  2(3-\nu)	\\
 \end{Bmatrix}
\chi_0^{2n}
\, .		\label{eq:weak}
\end{equation}
Similarly as in \Eqref{eq:weakp} above, these weak field expansions start with terms $\sim(e{\cal E}_0)^2$ and contain only even powers of $e{\cal E}_0$.
Correspondingly, and in line with expectations, for the real part of the polarization tensor in the weak field limit in Eqs.~\eqref{eq:weakT} and \eqref{eq:weak} the only effect of the inhomogeneity is a $d$ dependent modification of the expansion coefficients
We emphasize that the integrations over $\nu$ in Eqs.~\eqref{eq:weakT} and \eqref{eq:weak} could in principle be carried out and be expressed in terms of hypergeometric functions. As this does not come with any new insights we prefer to leave it unperformed for general kinematics. 
On the other hand, for on-shell photons with $k^2=0$ we have $\phi_0\to1$ and the integrations over $\nu$ simplify significantly.
In this limit, they can be readily evaluated and yield the following compact expressions,
\begin{equation}
    \pi_{T}^d\big|_{k^2=0} = -\frac{2\alpha}{\sqrt{\pi}}\sum_{n=1}^{\infty}\frac{1}{4}\frac{(3n-1)(3n-1)!\left[\Gamma(2n-\frac{d}{2})+\frac{d}{2}\Gamma(2n-1-\frac{d}{2})\right]}{48^{n}\Gamma(2n+\frac{5}{2}) \Gamma(n)}\,\chi_0^{2n}
\label{eq:crossedweakT}
\end{equation}
and
\begin{equation}
    \begin{Bmatrix}
     \tilde\pi_{FF}^d \\
     \tilde\pi^d_{{}^\star\!F{}^\star\!F}
    \end{Bmatrix}\bigg|_{k^2=0}
    =-  \frac{2\alpha}{\sqrt{\pi}}m^2\sum_{n=1}^{\infty}
    \frac{(3n-2)!\Gamma(2n-\frac{d}{2})}{48^{n}\Gamma(2n+\frac{3}{2})\Gamma(n)}
    \begin{Bmatrix}
    3n+1\\
    6n+1
    \end{Bmatrix}
    \chi_0^{2n}\,.
\label{eq:crossedweakF}
\end{equation}
The perturbative weak field expansions for $k^2=0$ given in Eqs.~\eqref{eq:crossedweakT} and \eqref{eq:crossedweakF} clearly do not feature any imaginary parts.
However, even for $k^2=0$ the components of the photon polarization tensor in a crossed field are known to generically feature manifestly non-perturbative imaginary parts.
With the help of the asymptotic expansion for ${\rm Hi}(z)$ given in formula 9.12.29 of Ref.~\cite{NIST} that also accounts for a contribution that cannot be exclusively represented in terms of a power series in $1/z$, the latter can be easily extracted from Eqs.~\eqref{eq:crossedTHi} and \eqref{eq:crossedFHi}.
Using formula 9.12.29 of Ref.~\cite{NIST} for ${\rm Hi}(z)$ and determining the analogous representations for ${\rm Hi}'(z)$ and ${\rm Hi}''(z)$ therefrom by differentiation, one finds that precisely the non-power series contributions constitute the imaginary parts of Eqs.~\eqref{eq:crossedTHi} and \eqref{eq:crossedFHi}; 
to this end note that ${\rm e}^{{\rm i}\frac{\pi}{3}}\rho^{-1/4}\sim{\rm e}^{-{\rm i}\frac{\pi}{3}}\rho^{1/4}\sim\rho^{3/4}\sim{\rm i}$ while $(\rho/\phi_0)^{3/2}$ is real-valued.
Upon limiting ourselves to the most relevant limit of on-shell photons with $k^2=0$, we first substitute $(1+t)/(1-\nu^2)\to t$ and make use of the fact that the integration domain of the resulting double integral can be re-expressed as $\int_0^1{\rm d}\nu\int_{1/(1-\nu^2)}^\infty{\rm d}t=\int_1^\infty{\rm d}t\int_0^{\sqrt{1-1/t}}{\rm d}\nu$. With the additional substitution $t(1-\nu^2)\to\nu$ we then obtain
\begin{multline}
  {\rm Im}\bigl\{\pi_T^d\bigr\}\big|_{k^2=0} = \frac{\alpha}{3\sqrt{6\pi}} \frac{1}{\Gamma(\frac{d}{2})}\sum_{n=0}^\infty c_n(-1)^n \\
  \times\int_1^\infty\frac{{\rm d}t}{t}\int_1^t \mathrm{d}\nu\, \Bigl(1-\frac{\nu}{t}\Bigr)^{\frac{1}{2}}\Bigl(2+\frac{\nu}{t}\Bigr)\frac{(\nu-1)^{\frac{d}{2}}}{\nu^{\frac{2}{3}}}\Bigl(\frac{1}{\nu-1}+\frac{d}{2}\Bigr)\Bigl(\frac{3}{8}\frac{\chi_0}{t}\Bigr)^{n-\frac{1}{2}}\,{\rm e}^{-\frac{8}{3}\frac{t}{\chi_0}}
  \label{eq:ImcrossedTHi}
 \end{multline}
and
\begin{equation}
    {\rm Im}\begin{Bmatrix}
     \tilde\pi_{FF}^d \\
     \tilde\pi^d_{{}^\star\!F{}^\star\!F}
 \end{Bmatrix}\bigg|_{k^2=0}
 = -\frac{\alpha}{3\sqrt{6}}m^2\frac{1}{\Gamma(\frac{3+d}{2})}\sum_{n=0}^\infty \tilde{c}_n(-1)^n \int_1^\infty{\rm d}t 
 \begin{Bmatrix}
   (4+3d)t-1   \\
   (4+6d)t+2  
\end{Bmatrix} \frac{(t-1)^{\frac{d-1}{2}}}{t^{\frac{3}{2}}}\Bigl(\frac{3}{8}\frac{\chi_0}{t}\Bigr)^{n+\frac{1}{2}}\,{\rm e}^{-\frac{8}{3}\frac{t}{\chi_0}}\,.
 \label{eq:ImcrossedFHi}
\end{equation}
Here, the expansion coefficients are given by
\begin{equation}
\begin{split}
    &c_0=1,\quad c_1=u_1\quad\text{and}\quad c_n=u_n-2(n-1)u_{n-1}+\Bigl[\frac{5}{36}+(n-1)(n-2)\Bigr]u_{n-2}\quad\text{for}\quad n\geq2,\\
    &\tilde{c}_0=1\quad\text{and}\quad \tilde{c}_n=u_n+\Bigl(\frac{5}{6}-n\Bigr)u_{n-1}\quad\text{for}\quad n\geq1,
    \end{split}
\end{equation}
with (Ref.~\cite{NIST}: 9.7.1)
\begin{equation}
    u_0=1\quad\text{and}\quad u_n=\frac{(2n+1)(2n+3)(2n+5)\cdots(6n-1)}{216^n n!} \quad\text{for}\quad n\geq1\,.
\end{equation}
We emphasize that in \Eqref{eq:ImcrossedFHi} the integration over $\nu$ could even be carried out explicitly in terms of elementary functions, leaving us with a single parameter integral to be performed.
The resulting expression~\eqref{eq:ImcrossedFHi} is even valid for the case of $d=0$ and thus holds for field amplitude profiles~\eqref{eq:calE} with $0\leq d\leq 3$ inhomogeneous directions.
In fact, also the integration over $t$ in \Eqref{eq:ImcrossedFHi} can be taken and expressed in terms of the Whittaker hypergeometric function via formula 3.383.4 of Ref.~\cite{Gradshteyn}, which then can be expanded for $\chi_0\ll1$ with the help of Ref.~\cite{Gradshteyn}: 9.383.4. This yields the compact expression, 
\begin{equation}
    {\rm Im}\begin{Bmatrix}
     \tilde\pi_{FF}^d \\
     \tilde\pi^d_{{}^\star\!F{}^\star\!F}
 \end{Bmatrix}\bigg|_{k^2=0}
 = -\sqrt{\frac{2}{3}}\,\alpha m^2
 \begin{Bmatrix}
   1   \\
   2  
\end{Bmatrix} \Bigl(\frac{3}{8}\chi_0\Bigr)^{1+\frac{d}{2}}{\rm e}^{-\frac{8}{3}\frac{1}{\chi_0}}\bigl[1+{\cal O}(\chi_0)\bigr]\,,
 \label{eq:ImcrossedFHiLO}
\end{equation}
that clearly recovers the well-known $d=0$ result.
Alternatively the result in \Eqref{eq:ImcrossedFHiLO} can be readily obtained by performing a Taylor expansion of the prefactor of the exponential function in the integrand of \Eqref{eq:ImcrossedFHi} about $t=1$ prior to carrying out the integration; this is justified because the strong exponential damping with $t$ for $\chi_0\ll1$ ensures that the main contribution of the integral over $t$ arises from the vicinity of its lower bound.
A similar strategy can be adopted to extract the leading contribution to \Eqref{eq:ImcrossedTHi} for $\chi_0\ll1$: To this end, we first expand the integrand of the double integral in \Eqref{eq:ImcrossedTHi} about $\nu=t$ to leading order, i.e., ${\cal O}((t-\nu)^{1/2})$ and perform the integration over $\nu$. It is immediately obvious that higher order contributions come with additional powers of $t-1$ and thus only give rise to subleading terms in the subsequent integration over $t$. The latter one is carried out along the lines just discussed as an alternative route to \Eqref{eq:ImcrossedFHiLO}. This results in
\begin{equation}
  {\rm Im}\bigl\{\pi_T^d\bigr\}\big|_{k^2=0} = \sqrt{\frac{2}{3}}\frac{\alpha}{3\sqrt{\pi}} \frac{\Gamma(\frac{3+d}{2})}{\Gamma(\frac{d}{2})} \Bigl(\frac{3}{8}\chi_0\Bigr)^{1+\frac{d}{2}} {\rm e}^{-\frac{8}{3}\frac{1}{\chi_0}}\bigl[1+{\cal O}(\chi_0)\bigr]\,.
  \label{eq:ImcrossedTHiLO}
 \end{equation}
Note that $\Gamma(\frac{3+d}{2})/\Gamma(\frac{d}{2})\big|_{d=0}=0$, such that for $\chi_0\ll1$ and $k^2=0$ 
there is no contribution scaling as $\sim\chi_0$ to
the imaginary part of $\pi_T$ in a uniform constant field.
From Eqs.~\eqref{eq:ImcrossedFHiLO} and \Eqref{eq:ImcrossedTHiLO} we infer that, as opposed to its real part, the imaginary part of the polarization tensor for weak crossed fields evaluated at $k^2=0$ shows a pronounced dependence on the number of inhomogeneous directions of the field inhomogeneity~\eqref{eq:calE}. While the overall non-perturbative exponential suppression remains unaffected by the dimension $d$ of the inhomogeneity, in essence the imaginary part gets additionally suppressed by a factor of $(3\chi_0/8)^{1/2}\ll1$ for each increase of $d\to d+1$.

Finally, we turn to the strong field limit $\chi_0\gg1$ of the polarization tensor in the special case of $k^2=0$.
For this analysis it is helpful to first rewrite Eqs.~\eqref{eq:crossedTHi} and \eqref{eq:crossedFHi}
specialized to $k^2=0$, using the substitutions introduced right above \Eqref{eq:ImcrossedTHi}.
However, after having implemented these substitutions, here we additionally rescale the integration variable $t/\chi_0\to t$ such as to render $\rho_0$ independent of $\chi_0$.
Therewith, we obtain
\begin{equation}
    \pi_T^d\big|_{k^2=0}=-\frac{\alpha}{9}\chi_0^{\frac{1}{3}+\frac{d}{2}}\frac{1}{\Gamma(\frac{d}{2})}\int_{\frac{1}{\chi_0}}^\infty\frac{\mathrm{d}t}{t}\,\mathrm{Hi}''(\rho_0)\int_{\frac{1}{\chi_0}}^t\mathrm{d}\nu\,\frac{\bigl(1-\frac{\nu}{t}\bigr)^{\frac{1}{2}}\bigl(2+\frac{\nu}{t}\bigr)\bigl(\nu-\frac{1}{\chi_0}\bigr)^{\frac{d}{2}-1}}{\nu^\frac{2}{3}}\biggl\{\frac{1}{\chi_0}+\frac{d}{2}\Bigl(\nu-\frac{1}{\chi_0}\Bigr)\biggr\}
    \label{eq:crossedTHik20}
\end{equation}
and
\begin{equation}
    \begin{Bmatrix}
     \tilde\pi_{FF}^d \vspace*{1mm}\\
     \tilde\pi^d_{{}^\star\!F{}^\star\!F}
 \end{Bmatrix}\bigg|_{k^2=0}
 = -\frac{\sqrt{\pi}\alpha}{6}\,m^2  \chi_0^{\frac{d}{2}} \frac{1}{\Gamma(\frac{3+d}{2})}\int_{\frac{1}{\chi_0}}^\infty \frac{\mathrm{d}t}{t}\,\frac{\bigl(t-\frac{1}{\chi_0}\bigr)^{\frac{d-1}{2}}}{t^{\frac{1}{2}}}\,\frac{\mathrm{Hi}'(\rho_0)}{\rho_0}
 \begin{Bmatrix}
     (4+3d)t-\frac{1}{\chi_0} \vspace*{1mm}\\ 
     (4+6d)t+\frac{2}{\chi_0}
 \end{Bmatrix},
 \label{eq:crossedFHik20}
\end{equation}
with
\begin{equation}
\rho_0=\mathrm{e}^{-\mathrm{i}\frac{2\pi}{3}}(4t)^{\frac{2}{3}}\,. \label{eq:rho0}
\end{equation}
Similarly as for the imaginary part in \Eqref{eq:ImcrossedFHi} above, the integration over $\nu$ in \Eqref{eq:crossedFHik20} could once again be performed explicitly in terms of elementary functions, and the resulting expression also holds for $d=0$.
We remark that the integral over $\nu$ in \Eqref{eq:crossedTHik20} can also be carried out and expressed via hypergeometric functions. However, as this representation is not particularly instructive, we refrain from giving it here.

Because their representation is considerably simpler, we first extract the leading strong field behavior of $\tilde\pi_{FF}^d$ and $\tilde\pi^d_{{}^\star\!F{}^\star\!F}$.
To this end, in a first step we introduce an auxiliary parameter $c$ fulfilling $1/\chi_0\ll c\ll1$ to split the integral in \Eqref{eq:crossedFHik20} into the two domains $1/\chi_0\leq t\leq c$ and $c<t\leq\infty$. By using the Maclaurin series of ${\rm Hi}'(\rho_0)$ (Ref.~\cite{NIST}: 9.12.18) in the former, it is easy to verify that its leading contribution for $\chi_0\gg1$ scales as $\chi_0^{2/3}$. At the same time, the leading contribution of the latter clearly scales as $\chi_0^{d/2}$. In turn, we have established that the strong field limit of \Eqref{eq:crossedFHik20} scales as $\chi_0^{2/3}$ for $d<4/3$ and as $\chi_0^{d/2}$ for $d>4/3$.
Upon replacing the derivative of Scorer's function in \Eqref{eq:crossedFHik20} by its leading term ${\rm Hi}'(\rho_0)\to\Gamma(2/3)/(3^{1/3}\pi)$, for $d<4/3$ the integral over the full interval $1/\chi_0\leq t\leq\infty$ remains finite and can be evaluated in terms of elementary functions. This yields the following explicit result for the relevant contribution $\sim\chi_0^{2/3}$,
\begin{equation}
    \begin{Bmatrix}
     \tilde\pi_{FF}^d \\
     \tilde\pi^d_{{}^\star\!F{}^\star\!F}
 \end{Bmatrix}\bigg|_{k^2=0}
 \xrightarrow{\chi_0\gg1}\ \alpha\,m^2\chi_0^{\frac{2}{3}}\frac{6^{\frac{2}{3}}\Gamma(\frac{2}{3})\Gamma(\frac{2}{3}-\frac{d}{2})}{7\sqrt{\pi}\,\Gamma(\frac{1}{6})}\,2
 \mathrm{e}^{-\mathrm{i}\frac{\pi}{3}}
 \begin{Bmatrix}
     2 \\ 
     3
 \end{Bmatrix} \quad\text{for}\quad d<4/3\,.
 \label{eq:crossedFHik20sf1}
\end{equation}
On the other hand, the contribution $\sim\chi_0^{d/2}$ for $d>4/3$ follows from \Eqref{eq:crossedFHik20} by taking the limit $1/\chi_0\to0$ in the integral, which also yields a manifestly finite expression. Correspondingly, we have
\begin{equation}
    \begin{split}
    \begin{Bmatrix}
     \tilde\pi_{FF}^d \\
     \tilde\pi^d_{{}^\star\!F{}^\star\!F}
 \end{Bmatrix}\bigg|_{k^2=0}
  \xrightarrow{\chi_0\gg1}\ &-\frac{\sqrt{\pi}\alpha}{6}\,m^2  \chi_0^{\frac{d}{2}} \frac{1}{\Gamma(\frac{3+d}{2})}\int_0^\infty \frac{\mathrm{d}t}{t}\,t^{\frac{d}{2}}\,\frac{\mathrm{Hi}'(\rho_0)}{\rho_0}
 \begin{Bmatrix}
     4+3d \\ 
     4+6d
 \end{Bmatrix} \\
  =\ &\,\alpha\,m^2\chi_0^{\frac{d}{2}} \,\frac{\Gamma(\frac{3d}{4}-1)\Gamma(1-\frac{d}{4})}{4\sqrt{\pi}\,(4\sqrt{3})^{\frac{d}{2}}\,\Gamma(\frac{3+d}{2})}\,{\rm e}^{-{\rm i}\frac{\pi}{4}d}
 \begin{Bmatrix}
     4+3d \\ 
     4+6d
 \end{Bmatrix}\quad\text{for}\quad d>4/3\,,
 \label{eq:crossedFHik20sf2}
 \end{split}
\end{equation}
where the integral over $t$ could be straightforwardly evaluated analytically by expressing the derivative of Scorer's function in terms of its integral representation via \Eqref{eq:Hi}; upon performing the integration over $t$, the integration over $s$ can be easily taken. 
In summary, we showed that the strong field limits of $\tilde\pi_{FF}^d$ and $\tilde\pi^d_{{}^\star\!F{}^\star\!F}$ scale as $\sim\chi_0^{2/3}$ in both constant crossed fields and crossed fields with a single inhomogeneous direction along which the field features a Lorentzian amplitude profile~\eqref{eq:calE}. For Lorentzian crossed field inhomogeneities with $2\leq d\leq3$ inhomogeneous directions the strong field scaling is found to be enhanced to $\sim\chi_0^{d/2}$ and thus features an explicit dependence on the dimension of the inhomogeneity. We also determined the associated numerical coefficients depending on $d$.  

Next, we aim at a similar analysis for the strong field limit of $\pi_T^d$. To this end, we first note that with the substitution $1/(1-\nu^2)/\chi_0\to t$ the result for $d=0$ that follows with the replacements detailed below \Eqref{eq:rho} from \Eqref{eq:crossedTHi} can be compactly represented as
\begin{equation}
\begin{split}
  \pi_T^{d=0}\big|_{k^2=0} =\ & -\frac{2\alpha}{9} \int_{\frac{1}{\chi_0}}^\infty \frac{\mathrm{d}t}{t}\,\frac{\bigl(t-\frac{1}{\chi_0}\bigr)^{\frac{1}{2}}\bigl(t+\frac{1}{2\chi_0}\bigr)}{t^{\frac{3}{2}}} \,\mathrm{Hi}''(\rho_0) \\
  \xrightarrow{\chi_0\gg1}\ &
  -\frac{2}{9}\frac{\alpha}{\pi}\ln\chi_0 \,,
  \label{eq:crossedTHik20sf0}
 \end{split}
\end{equation}
where the first line still holds for generic values of $\chi_0$. The limit provided in the second line of \Eqref{eq:crossedTHik20sf0} can be straightforwardly extracted by using the auxiliary parameter $c$ introduced in the paragraph below \Eqref{eq:rho0} and employing the Mclaurin series of ${\rm Hi}''(\rho_0)$ in the relevant domain $1/\chi_0\leq t\leq c$; the latter follows from formula 9.12.18 of Ref.~\cite{NIST} by differentiation.
Equation~\eqref{eq:crossedTHik20sf0} implies that, similar to its magnetic/electric field analogue in \Eqref{eq:forwardstrong0D}, the leading strong field behavior of $\pi_T^{d=0}$ is characterized by a logarithmic scaling: in the present case we obviously have $\lim_{\chi_0\to\infty}\pi_T^{d=0}=-\alpha\beta_1\ln\chi_0^{2/3}$ \cite{Mironov:2020gbi}.

On the other hand, applying the same strategy as used for the determination of the strong field scaling of $\tilde\pi_{FF}^d$ and $\tilde\pi^d_{{}^\star\!F{}^\star\!F}$ above, \Eqref{eq:crossedTHik20} allows us to establish that the strong field limit of $\pi_T^d$ scales as $\chi_0^{1/3+d/2}$ for $d\geq1$.
In turn, for $d\geq1$ the leading strong field behavior of $\pi_T^d$ can be obtained form \Eqref{eq:crossedTHik20} by once again taking the limit $1/\chi_0\to0$ in the integral, yielding
\begin{equation}
\begin{split}
\pi_T^d\big|_{k^2=0}\xrightarrow{\chi_0\gg1}\ &-\frac{\alpha}{9}\chi_0^{\frac{1}{3}+\frac{d}{2}}\frac{\frac{d}{2}}{\Gamma(\frac{d}{2})}\int_0^\infty\frac{\mathrm{d}t}{t}\,\mathrm{Hi}''(\rho_0)\int_0^t\mathrm{d}\nu\,\Bigl(1-\frac{\nu}{t}\Bigr)^{\frac{1}{2}}\Bigl(2+\frac{\nu}{t}\Bigr)\,\nu^{\frac{d}{2}-\frac{2}{3}} \\
    =\ &\,-\alpha\,\chi_0^{\frac{1}{3}+\frac{d}{2}}\frac{3d}{2+3d}\frac{\Gamma(\frac{7}{3}+\frac{d}{2})\Gamma(\frac{1}{2}+\frac{3d}{4})\Gamma(\frac{5}{6}-\frac{d}{4})}{4\sqrt{\pi}\,(4\sqrt{3})^{\frac{1}{3}+\frac{d}{2}}\,\Gamma(\frac{d}{2})\Gamma(\frac{17}{6}+\frac{d}{2})}\,{\rm e}^{-{\rm i}\frac{\pi}{4}(\frac{2}{3}+d)}
    \,.
    \label{eq:crossedTHik20sf}
    \end{split}
\end{equation}
While the integration over $\nu$ in \Eqref{eq:crossedTHik20sf} could be readily taken and be expressed in terms of gamma functions, to perform the $t$ integration we used the same strategy as in \Eqref{eq:crossedFHik20sf2} and re-expressed the second derivative of Scorer's function in terms of its integral representation via \Eqref{eq:Hi} in an intermediate step.
A comparison of Eqs.~\eqref{eq:crossedFHik20sf1}-\eqref{eq:crossedFHik20sf2} and \eqref{eq:crossedTHik20sf0}-\eqref{eq:crossedTHik20sf} unveils that whereas $\tilde\pi_{FF}^d$ and $\tilde\pi^d_{{}^\star\!F{}^\star\!F}$ dominate over $\pi_{T}^d$ in the strong field limit for $d=0$, interestingly this behavior is reversed for $d\geq1$ where the dominant contribution arises from $\pi_{T}^d\sim\chi_0^{(2+3d)/6}$.

\subsection{Physical implications}\label{sec:phys}

The results for the photon polarization tensor in an electromagnetic field inhomogeneity with Lorentzian amplitude profile~\eqref{eq:calE} derived in Sec.~\ref{sec:poltensor} have several direct consequences for observables to be studied in quantum vacuum experiments.
To be specific, in this section we only feature a selection of effects that can be reliably studied within the limitations of both approaches (i) and (ii) invoked in in Sec.~\ref{sec:poltensor}.

In general, the real part of the polarization tensor encodes dispersive effects on photon propagation and its imaginary part absorptive effects. A prominent example of the former are different polarization eigenmodes that are effectively imprinted onto the vacuum by the background field. These generically come with distinct indices of refraction for probe photons and lead, e.g., to the experimental signature of vacuum birefringence. On the other hand, a well-known absorptive effect is the background-field-assisted conversion of probe photons into real electron-positron pairs inherently coming with a loss of probe photons. 

One can easily convince oneself that the normalized four-vectors $\epsilon_{(1)}^\mu(k):=(kF)^\mu/\sqrt{(kF)^2}$ and $\epsilon_{(2)}^\mu(k):=(k{}^\star\!F)^\mu/\sqrt{(kF)^2}$ form a basis for the photon polarizations transverse to $k^\mu$ for $k^2=0$, where $(k{}^\star\!F)^2=(kF)^2$; cf., e.g., Ref.~\cite{Bialynicka-Birula:1970nlh}.
These vectors are independent of the amplitude profile of the background field by construction and depend only on its direction $\hat{\vec{\kappa}}$. They fulfill $\epsilon_{(p)\mu}(k)\epsilon_{(p)}^\mu(k)=1$ and $k_\mu\epsilon_{(1
)}^\mu(k)=k_\mu\epsilon_{(2)}^\mu(k)=\epsilon_{(1)\mu}(k)\epsilon_{(2)}^\mu(k)=0$. 
In line with that, they also span the physical transverse photon polarizations at zero background field. The polarization mode characterized by $\epsilon_{(1)}^\mu(k)$ is polarized perpendicular to the one associated with $\epsilon_{(2)}^\mu(k)$.
Hence, a generic transverse monochromatic positive energy photon field solving the linear Maxwell equations can be represented as
\begin{equation}
 a^\mu(k)=\frac{1}{2}\bigl\{\mathfrak{a}_1\,\epsilon_{(1)}^\mu(k)+\mathfrak{a}_2\,\epsilon_{(2)}^\mu(k)\bigr\}{\rm e}^{{\rm i}kx}\Big|_{k^2=0}\,, \label{eq:genE>0probePW}
\end{equation}
with amplitudes $\mathfrak{a}_{1,2}$, that are real (complex) valued for linear (elliptic) polarizations. 
Clearly, also the above expressions for the photon polarization tensor in a purely magnetic or electric field in Eqs.~\eqref{eq:PiF}, \eqref{eq:PiFd} and a crossed field in Eqs.~\eqref{eq:PiCrossed}, \eqref{eq:PiFcrossd} are already written in this basis.
By definition, a monochromatic plane wave is infinitely extended in the directions perpendicular to its propagation direction $\vec{k}$ and lasts infinitely long.
Correspondingly, it always samples the full extent of the field inhomogeneity transverse to $\vec{k}$.

Specifically in the strict forward limit, where $k'^\mu=k^\mu$, the polarization tensor is diagonal in this basis and the only non-vanishing polarization matrix elements for the crossed field case are
\begin{equation}
\begin{Bmatrix}
     \epsilon_{(1)}^\mu(k)\,\Pi_{\mu\nu}(-k,k)\,\epsilon_{(1)}^\nu(k) \\
     \epsilon_{(2)}^\mu(k)\,\Pi_{\mu\nu}(-k,k)\,\epsilon_{(2)}^\nu(k)
 \end{Bmatrix}\bigg|_{k^2=0}
    =(2\pi)^{4-d}\delta^{(4-d)}(k=0)\biggl(\,\prod_{i=1}^d w_i\biggr)\Bigl(\frac{\pi}{4}\Bigr)^{\frac{d}{2}}
    \begin{Bmatrix}
     \tilde\pi_{FF}^d \\
     \tilde\pi^d_{{}^\star\!F{}^\star\!F}
 \end{Bmatrix}\bigg|_{k^2=0}\,.
 \label{eq:ePIeCrossed}
\end{equation}
We emphasize that while clearly only the components of the polarization tensor labeled by $p\in\{FF,{}^\star\!F{}^\star\!F\}$ exhibit a direct coupling to real transverse on-shell photons the study of the $p=T$ component in Sec.~\ref{sec:poltensor} is also very relevant because it encodes important information about the renormalization group properties of the theory; see also the corresponding comment in the paragraph below \Eqref{eq:pi_sFsF}.

Apart from an overall normalization, the expressions in \Eqref{eq:ePIeCrossed} amount to the background field dependent asymptotic forward scattering amplitudes $T_{\epsilon'\epsilon}(k)$ \cite{Dinu:2013gaa} for probe photons of wave vector $k^\mu=\omega(1,\vec{k}/|\vec{k}|)$ in the polarization basis spanned by $\epsilon_{(1)}^\mu$ and $\epsilon_{(2)}^\mu$.
To be precise, in our conventions we have \cite{Dinu:2013gaa,Karbstein:2015xra}
\begin{equation}
    T_{\epsilon'\epsilon}(k)=-\frac{\epsilon'^\mu(k)\,\Pi_{\mu\nu}(-k,k)\,\epsilon^\nu(k)}{2\omega V_\perp^{(3)}}\bigg|_{k^2=0}\,. \label{eq:Te'e}
\end{equation}
Here, $V_\perp^{(3)}$ denotes the three-dimensional space-time volume perpendicular to $\vec{k}$; $\epsilon'^\mu(k)$ and $\epsilon^\mu(k)$ are normalized polarization four-vectors transverse to $k^\mu$.
Also recall that the overall factor of $(2\pi)^{4-d}\delta^{(4-d)}(k=0)$ in \Eqref{eq:ePIeCrossed} amounts to the $(4-d)$-dimensional space-time volume $V^{(4-d)}$ associated with the homogeneous directions of the background field; cf. the paragraph below \Eqref{eq:fkks}.
While the real part of \Eqref{eq:Te'e} can be attributed to dispersive effects, its imaginary part describes an photon-absorbing property of the quantum vacuum that can be related to the effect of electron positron pair-production.
A comparison of Eqs.~\eqref{eq:PiF} and \eqref{eq:PiCrossed} unveils that the analogous results for a purely magnetic/electric field pointing in direction $\hat{\vec{\kappa}}$ can be obtained from~\Eqref{eq:ePIeCrossed} by substituting 
$\tilde\pi_p^d\to(kF)^2/(2{\cal F})\,\pi_p^d={\rm sign}({\cal F})\,(\vec{k}\times\hat{\vec{\kappa}})^2\,\pi_p^d$ with $p\in\{FF,{}^\star\!F{}^\star\!F\}$, where the latter identity holds only for $k^2=0$. Recall that in our conventions ${\rm sign}({\cal F})$ is positive (negative) for the purely magnetic (electric) field case.

Using notations closely paralleling those of Ref.~\cite{Dinu:2013gaa}, the associated $\epsilon^\mu(k)\to\epsilon'^\mu(k)$ forward scattering probability follows from \Eqref{eq:Te'e} as 
\begin{equation}
    \mathbb{P}_{\epsilon'\epsilon}(k)=\bigl|T_{\epsilon'\epsilon}(k)\bigr|^2\,. \label{eq:Pe'e}
\end{equation}
A non-vanishing polarization-flip probability can be attributed to a birefringence property of the quantum vacuum in the presence of the background field \cite{Toll:1952rq,Erber:1962dsd,Klein:1964}. 
On the other hand, the imaginary part of the non-flip $\epsilon^\mu(k)\to\epsilon^\mu(k)$ scattering amplitude is related to the total probability of electron-positron pair production induced by $\epsilon^\mu(k)$-polarized probe-photons of wave vector $k^\mu=\omega(1,\vec{k}/|\vec{k}|)$  via the optical theorem (cf., e.g., \cite{Nikishov:1964zza,Tsai:1974fa}) as 
\begin{equation}
 \mathbb{P}_\epsilon({\rm pair})=2\,{\rm Im}\bigl\{T_{\epsilon\epsilon}(k)\bigr\}\,. \label{eq:Ppair}
\end{equation}

In the linearly polarized case it is particularly convenient to represent \Eqref{eq:genE>0probePW} in terms of a single amplitude $\mathfrak{a}$ and an angle $\beta$ that parameterizes the possible linear polarizations. To this end one identifies $\mathfrak{a}_1=\mathfrak{a}\cos\beta$ and $\mathfrak{a}_2=\mathfrak{a}\sin\beta$.
The associated polarization vector is then given by $\epsilon^\mu(k)=\epsilon_{(1)}^\mu(k)\cos\beta+\epsilon_{(2)}^\mu(k)\sin\beta$, and that for the perpendicularly polarized mode by $\epsilon_\perp^\mu(k)=-\epsilon_{(1)}^\mu(k)\sin\beta+\epsilon_{(2)}^\mu(k)\cos\beta$.
In this basis we generically encounter both background field dependent polarization flip and no-flip amplitudes. These are easily inferred from Eqs.~\eqref{eq:ePIeCrossed} and \eqref{eq:Te'e}.
The former is given by
\begin{equation}
    T_{\epsilon_\perp\epsilon}(k)=\frac{V^{(4-d)}}{V_\perp^{(3)}}\biggl(\,\prod_{i=1}^d w_i\biggr)\Bigl(\frac{\pi}{4}\Bigr)^{\frac{d}{2}}\,\frac{\sin(2\beta)}{2}\frac{\tilde\pi_{FF}^d-\tilde\pi^d_{{}^\star\!F{}^\star\!F}}{2\omega}\bigg|_{k^2=0}\,,
 \label{eq:ePIeCrossed_flip}
\end{equation}
and the latter by
\begin{equation}
    T_{\epsilon\epsilon}(k)=-\frac{V^{(4-d)}}{V_\perp^{(3)}}\biggl(\,\prod_{i=1}^d w_i\biggr)\Bigl(\frac{\pi}{4}\Bigr)^{\frac{d}{2}}\,\frac{\tilde\pi_{FF}^d\cos^2\beta+\tilde\pi^d_{{}^\star\!F{}^\star\!F}\sin^2\beta}{2\omega}\bigg|_{k^2=0}\,.
 \label{eq:ePIeCrossed_noflip}
\end{equation}

While these expressions can be analyzed for arbitrary probe-photon propagation directions $\vec{k}$, for simplicity it is most convenient to assume that $\vec{k}$ is aligned with one of the coordinate axes.
For directions perpendicular to $\vec{k}$ in which the background field~\eqref{eq:calE} is homogeneous the corresponding extents in the ratio $V^{(4-d)}/V^{(3)}_\perp$ cancel out.
Conversely, if the background field is inhomogeneous in such a direction we are left with a factor of $w_i/L_i$ from the ratio $(\,\prod_{i=1}^d w_i)/V^{(3)}_\perp$. Here, $L_i\gg w_i$ measures the -- formally infinite -- extent of the probe in $i$ direction.
Similarly, a factor of $L_i$ from $V^{(4-d)}$ is retained if $\vec{k}$ points in a homogeneous direction $i$, and a factor of $w_i$ out of $(\,\prod_{i=1}^d w_i)$ for an inhomogeneous propagation direction.

Subsequently, we provide explicit results for the polarization-flip probability $\mathbb{P}_{\epsilon_\perp\epsilon}(k)$ in \Eqref{eq:Pe'e} for linearly polarized probe photons and the associated electron-positron pair production probability~\eqref{eq:Ppair}. Both of these quantities are -- at least in principle -- accessible in experiment. While the polarization-flip phenomenon in the weak field limit can be attributed to a dispersive property of the quantum vacuum subjected to the external field, in the strong field limit it typically arises from the combination of dispersive and absorptive effects.

First, we focus on the case of a crossed background field with amplitude profile~\eqref{eq:calE}.
We note that for on-shell photons as considered here and in the remainder of this article, the quantum nonlinear parameter of the peak field introduced in \Eqref{eq:chi0} can be recast into
\begin{equation}
    \chi_0=\frac{e{\cal E}_0}{m^2}\frac{\omega}{m}(1-\cos\theta_{\rm coll})\,,
    \label{eq:chi0_onshell}
\end{equation}
with collision angle $\theta_{\rm coll}$ defined as $\hat{\vec{\kappa}}\cdot\vec{k}/|\vec{k}|=\cos\theta_{\rm coll}$.
The polarization-flip probability for linearly polarized probe photons in weak crossed fields fulfilling $\chi_0\ll1$ can be obtained straightforwardly by plugging the leading order result in \Eqref{eq:crossedweakF} into Eqs.~\eqref{eq:Pe'e} and \eqref{eq:ePIeCrossed_flip}. This yields
\begin{equation}
    \mathbb{P}_{\epsilon_\perp\epsilon}(k)=\biggl(\frac{V^{(4-d)}}{V_\perp^{(3)}\lambdabar_{\rm C}}\,\prod_{i=1}^d w_i\biggr)^2\Bigl(\frac{\pi}{4}\Bigr)^d\,\sin^2(2\beta)\,
    \frac{\Gamma^2(2-\frac{d}{2})}{3600\pi^2}\,\alpha^2\Bigl(\frac{m}{\omega}\Bigr)^2
    \chi_0^{4}\,\bigl[1+{\cal O}(\chi_0^2)\bigr] \label{eq:Peperpe_smallchi0}
\end{equation}
for $0\leq d\leq3$.
The results in the complementary parameter regime of $\chi_0\gg1$ follow from
Eqs.~\eqref{eq:crossedFHik20sf1}, \eqref{eq:crossedFHik20sf2} and are
given by 
\begin{equation}
\begin{split}
    \mathbb{P}_{\epsilon_\perp\epsilon}(k)\big|_{d=0} \,&\xrightarrow{\chi_0\gg1}\ \biggl(\frac{V^{(4)}}{V_\perp^{(3)}\lambdabar_{\rm C}}\biggr)^2\,\sin^2(2\beta)\,\frac{6^{\frac{4}{3}}\Gamma^4(\frac{2}{3})\Gamma^2(\frac{5}{6})}{784\pi^3}\,\alpha^2\Bigl(\frac{m}{\omega}\Bigr)^2\chi_0^{\frac{4}{3}}\,,\\
    \mathbb{P}_{\epsilon_\perp\epsilon}(k)\big|_{d=1} \,&\xrightarrow{\chi_0\gg1}\ \biggl(\frac{V^{(3)}w_1}{V_\perp^{(3)}\lambdabar_{\rm C}}\biggr)^2\,\sin^2(2\beta)\,\frac{6^{\frac{4}{3}}\Gamma^2(\frac{2}{3})}{784}\,\alpha^2\Bigl(\frac{m}{\omega}\Bigr)^2\chi_0^{\frac{4}{3}}\,,\\
    \mathbb{P}_{\epsilon_\perp\epsilon}(k)\big|_{d=2} \,&\xrightarrow{\chi_0\gg1}\ \biggl(\frac{V^{(2)}w_1w_2}{V_\perp^{(3)}\lambdabar_{\rm C}}\biggr)^2\,\sin^2(2\beta)\,\frac{\pi^2}{3072}\,\alpha^2\Bigl(\frac{m}{\omega}\Bigr)^2\,\chi_0^2 \,,\\
    \mathbb{P}_{\epsilon_\perp\epsilon}(k)\big|_{d=3} \,&\xrightarrow{\chi_0\gg1}\ \biggl(\frac{V^{(1)}w_1w_2w_3}{V_\perp^{(3)}\lambdabar_{\rm C}}\biggr)^2\,\sin^2(2\beta)\,\frac{3^{\frac{5}{2}}\pi^2\Gamma^4(\frac{1}{4})}{2^{26}}\,\alpha^2\Bigl(\frac{m}{\omega}\Bigr)^2\chi_0^3 \,.
    \end{split} \label{eq:Peperpe_largechi0}
\end{equation}
Interestingly the scaling of the polarization-flip probability with $\chi_0$ remains the same for $0\leq d\leq1$, but changes and becomes more pronounced for larger values of $d$.

At the same time, Eqs.~\eqref{eq:ImcrossedFHiLO}, \eqref{eq:Ppair} and \eqref{eq:ePIeCrossed_noflip} imply that in the crossed field case the pair production probability in the weak field limit $\chi_0\ll1$ induced by linearly polarized probe photons can be expressed as
\begin{equation}
    \mathbb{P}_\epsilon({\rm pair})=\frac{V^{(4-d)}}{V_\perp^{(3)}\lambdabar_{\rm C}}\biggl(\,\prod_{i=1}^d w_i\biggr)(1-\cos\theta_{\rm coll})(\cos^2\beta+2\sin^2\beta)\,\frac{\sqrt{6}}{8}\,\alpha\,\frac{e{\cal E}_0}{m^2}
  \Bigl(\frac{\pi}{4}\frac{3}{8}\chi_0\Bigr)^{\frac{d}{2}}{\rm e}^{-\frac{8}{3}\frac{1}{\chi_0}}\bigl[1+{\cal O}(\chi_0)\bigr] \label{eq:Ppair_crossed}
\end{equation}
for $0\leq d\leq 3$. While the overall exponential suppression of the effect remains independent of $d$, \Eqref{eq:Ppair_crossed} clearly implies that the more pronounced the localization of the background field, the smaller the pair yield for fixed $\chi_0$.
At this point we emphasize that this result is also of interest and importance for studies of nonlinear Breit-Wheeler pair production \cite{Breit:1934zz,Reiss:1962nhe} in the collision of gamma rays with a focused intense laser beam; cf., e.g., \cite{Meuren:2014uia,Seipt:2020diz}. While laser beams are reasonably well-modeled as crossed fields, -- especially transverse to their propagation direction -- these typically feature field profiles different from Lorentzian ones. Nevertheless, similar localization effects are to be expected also there; cf., e.g., Refs.~\cite{Salgado:2021uua,Golub:2022cvd} and references therein.
On the other hand, from Eqs.~\eqref{eq:crossedFHik20sf1} and \eqref{eq:crossedFHik20sf2} we obtain the pair production probabilities in the the strong field limit,
\begin{equation}
    \begin{split}
    \mathbb{P}_\epsilon({\rm pair})\big|_{d=0}\,&\xrightarrow{\chi_0\gg1}\ \frac{V^{(4)}}{V_\perp^{(3)}\lambdabar_{\rm C}}\,(2\cos^2\beta+3\sin^2\beta)\,\frac{3^{\frac{5}{3}}\Gamma^4(\frac{2}{3})}{14\pi^2}\,\alpha\frac{m}{\omega}\chi_0^{\frac{2}{3}}\,, \\
    \mathbb{P}_\epsilon({\rm pair})\big|_{d=1}\,&\xrightarrow{\chi_0\gg1}\ \frac{V^{(3)}w_1}{V_\perp^{(3)}\lambdabar_{\rm C}}\,(2\cos^2\beta+3\sin^2\beta)\,\frac{6^{\frac{7}{6}}\Gamma(\frac{2}{3})}{14\sqrt{2}}\,\alpha\frac{m}{\omega}\chi_0^{\frac{2}{3}}\,, \\
    \mathbb{P}_\epsilon({\rm pair})\big|_{d=2}\,&\xrightarrow{\chi_0\gg1}\ \frac{V^{(2)}w_1w_2}{V_\perp^{(3)}\lambdabar_{\rm C}}\,(5\cos^2\beta+8\sin^2\beta)\,\frac{\pi\sqrt{3}}{72}\,\alpha\frac{m}{\omega}\chi_0\,, \\
    \mathbb{P}_\epsilon({\rm pair})\big|_{d=3}\,&\xrightarrow{\chi_0\gg1}\ \frac{V^{(1)}w_1w_2w_3}{V_\perp^{(3)}\lambdabar_{\rm C}}\,(13\cos^2\beta+22\sin^2\beta)\,\frac{3^{\frac{1}{4}}\pi\Gamma^2(\frac{1}{4})}{6144\sqrt{2}}\,\alpha\frac{m}{\omega}\chi_0^{\frac{3}{2}}\,.
    \end{split} \label{eq:eq:Ppair_crossed_strong}
\end{equation}
These expressions show a similar dependence on $d$ as \Eqref{eq:Peperpe_largechi0}.
In summary, we inferred that whereas in the perturbative limit the scaling of the polarization-flip probabilities~\eqref{eq:Peperpe_smallchi0} with $\chi_0$ remains unaltered relative to the reference case with $d=0$ and only the coefficients become $d$ dependent, the scaling of the non-perturbative results~\eqref{eq:Peperpe_largechi0}-\eqref{eq:eq:Ppair_crossed_strong} changes notably with $d$. 
Moreover, in passing we note that our results for the crossed field case with $\chi_0\gg1$ also touch upon questions relevant in the context of the Ritus-Narozhny conjecture \cite{Narozhnyi:1980dc,Fedotov:2016afw,Fedotov:2022ely}: They indicate that even in weakly localized crossed fields the strong field scaling at one loop may significantly deviate from the constant field ($d=0$) behavior $T_{\epsilon'\epsilon}\sim\alpha\chi_0^{2/3}$ at amplitude level.

Second, we discuss the analogous results for the purely magnetic or electric field case with the additional restriction on low-frequency $\omega/m\ll1$ probe photons.
The corresponding perturbative weak field $e{\cal E}_0/m^2\ll1$ result follows from Eqs.~\eqref{eq:weakp}, \eqref{eq:Pe'e} and \eqref{eq:ePIeCrossed_flip} and reads
\begin{equation}
    \mathbb{P}_{\epsilon_\perp\epsilon}(k)=\biggl(\frac{V^{(4-d)}}{V_\perp^{(3)}\lambda}\,\prod_{i=1}^d w_i\biggr)^2\Bigl(\frac{\pi}{4}\Bigr)^d\,\sin^4\theta_F\,\sin^2(2\beta)\,\frac{\Gamma^2(2-\tfrac{d}{2})}{900}\,\alpha^2\Bigl(\frac{e{\cal E}_0}{m^2}\Bigr)^4\,\bigl[1+{\cal O}\bigl((\tfrac{e{\cal E}_0}{m^2})^2\bigr)\bigr]\,. \label{eq:Peperpe_LOpert_B}
\end{equation}
Here, $\lambda=2\pi/\omega$ is the probe wavelength  and $\theta_F=\angle(\vec{k},\hat{\vec{\kappa}})$ the angle between the probe-photon propagation direction $\vec{k}$ and the direction $\hat{\vec{\kappa}}$ of the magnetic or electric field, respectively; cf. the corresponding discussion in the second paragraph below \Eqref{eq:calE}.
We note that the $d$ dependence of \Eqref{eq:Peperpe_LOpert_B} is exactly the same as that of \Eqref{eq:Peperpe_smallchi0}.
This is in line with expectations as in both cases the relevant contribution to the photon polarization tensor is quadratic in the coupling to the background field.
The results for the polarization-flip probability in the strong field limit $e{\cal E}_0/m^2\gg1$ following from Eqs.~\eqref{eq:forwardstrong0D}-\eqref{eq:forwardstrong3D} read
\begin{equation}
\begin{split}
    \mathbb{P}_{\epsilon_\perp\epsilon}(k)\big|_{d=0}\,&\xrightarrow{\frac{e{\cal E}_0}{m^2}\gg1}\ \biggl(\frac{V^{(4)}}{V_\perp^{(3)}\lambda}\biggr)^2\, \sin^4\theta_F\,\sin^2(2\beta)\frac{1}{36}\,\alpha^2\,\Bigl(\frac{e{\cal E}_0}{m^2}\Bigr)^2 \,, \\
    \mathbb{P}_{\epsilon_\perp\epsilon}(k)\big|_{d=1}\,&\xrightarrow{\frac{e{\cal E}_0}{m^2}\gg1}\ \biggl(\frac{V^{(3)}w_1}{V_\perp^{(3)}\lambda}\biggr)^2\, \sin^4\theta_F\,\sin^2(2\beta)\frac{\pi^2}{144}\,\alpha^2\,\Bigl(\frac{e{\cal E}_0}{m^2}\Bigr)^2\,,\\
    \mathbb{P}_{\epsilon_\perp\epsilon}(k)\big|_{d=2}\,&\xrightarrow{\frac{e{\cal E}_0}{m^2}\gg1}\ \biggl(\frac{V^{(2)}w_1w_2}{V_\perp^{(3)}\lambda}\biggr)^2\, \sin^4\theta_F\,\sin^2(2\beta)\frac{\pi^2}{576}\,\alpha^2\,\Bigl(\frac{e{\cal E}_0}{m^2}\Bigr)^2\ln^2\Bigl(\frac{e{\cal E}_0}{m^2}\Bigr)\,,\\
    \mathbb{P}_{\epsilon_\perp\epsilon}(k)\big|_{d=3}\,&\xrightarrow{\frac{e{\cal E}_0}{m^2}\gg1}\ \biggl(\frac{V^{(1)}w_1w_2w_3}{V_\perp^{(3)}\lambda}\biggr)^2\, \sin^4\theta_F\,\sin^2(2\beta)\frac{1}{512}\Bigl(\frac{\pi}{3}\zeta(\tfrac{3}{2})-\frac{7}{2\pi}\zeta(\tfrac{7}{2})\Bigr)^2\,\alpha^2\,\Bigl(\frac{e{\cal E}_0}{m^2}\Bigr)^3\,. \label{eq:Peperpe_strongB}
    \end{split}
\end{equation}
Similar as for the crossed-field case in \Eqref{eq:Peperpe_largechi0}, the scaling of \Eqref{eq:Peperpe_strongB} with the background field remains unchanged for $0\leq d\leq1$, but is enhanced for an increased number of inhomogeneous directions $2\leq d\leq3$. We emphasize that Eqs.~\eqref{eq:Peperpe_LOpert_B} and \eqref{eq:Peperpe_strongB} hold for both purely magnetic and electric background fields.

As detailed in Sec.~\ref{subsec:B+E}, in an electric field, but not in a purely magnetic field, the photon polarization tensor develops a non-vanishing imaginary part for on-shell probe photons in the low-frequency limit $\omega/m\ll1$.
From Eqs.~\eqref{eq:ImPipweakfield}, \eqref{eq:Ppair} and \eqref{eq:ePIeCrossed_noflip} we infer that in weak electric fields $e{\cal E}_0/m^2\ll1$ the associated total probability of electron-positron pair production stimulated by low-frequency photons can be expressed as
\begin{equation}
    \mathbb{P}_\epsilon({\rm pair})=\frac{V^{(4-d)}}{V_\perp^{(3)}\lambda}\biggl(\,\prod_{i=1}^d w_i\biggr)\,\sin^2\theta_F\cos^2\beta\,\frac{\pi}{16}\,\alpha\,\Bigl(\frac{1}{4}\frac{e{\cal E}_0}{m^2}\Bigr)^{\frac{d}{2}-2}\,\mathrm{e}^{-\frac{m^2}{e{\cal E}_0}\pi}\,
         \bigl[1+{\cal O}\bigl(\tfrac{e{\cal E}_0}{m^2}\bigr)\bigr]\,. \label{eq:Ppair_E}
\end{equation}
In line with its crossed field analogue~\eqref{eq:Ppair_crossed}, \Eqref{eq:Ppair_E} implies a reduction of the pair production probability with an increasing number of inhomogeneous directions $d$. 
The analogous result in the strong field limit $e{\cal E}_0/m^2\gg1$ readily follows from \Eqref{eq:ImSF_LO}, yielding
\begin{equation}
\begin{split}
    \mathbb{P}_\epsilon({\rm pair})\big|_{d=0}\,&\xrightarrow{\frac{e{\cal E}_0}{m^2}\gg1}\ \frac{V^{(4)}}{V_\perp^{(3)}\lambda}\sin^2\theta_F\sin^2\beta\,\frac{2}{3}\,\alpha\,\frac{e{\cal E}_0}{ m^2}\,, \\
    \mathbb{P}_\epsilon({\rm pair})\big|_{d=1}\,&\xrightarrow{\frac{e{\cal E}_0}{m^2}\gg1}\ \frac{V^{(3)}w_1}{V_\perp^{(3)}\lambda}\sin^2\theta_F\sin^2\beta\,\frac{\pi}{3}\,\alpha\,\frac{e{\cal E}_0}{ m^2}\,, \\
    \mathbb{P}_\epsilon({\rm pair})\big|_{d=2}\,&\xrightarrow{\frac{e{\cal E}_0}{m^2}\gg1}\ \frac{V^{(2)}w_1w_2}{V_\perp^{(3)}\lambda}\sin^2\theta_F\sin^2\beta\,\frac{\pi}{6}\,\alpha\,\frac{e{\cal E}_0}{ m^2}\, \ln\frac{e{\cal E}_0}{m^2}\,,  \\
    \mathbb{P}_\epsilon({\rm pair})\big|_{d=3}\,&\xrightarrow{\frac{e{\cal E}_0}{m^2}\gg1}\ \frac{V^{(1)}w_1w_2w_3}{V_\perp^{(3)}\lambda}\sin^2\theta_F\Bigl[21\zeta(\tfrac{7}{2})+\Bigl(\frac{8\pi^2}{3}\zeta(\tfrac{3}{2})-7\zeta(\tfrac{7}{2})\Bigr)\sin^2\beta\Bigr]\frac{1}{4^{\frac{5}{2}}\pi}\,\alpha\,\Bigl(\frac{e{\cal E}_0}{ m^2}\Bigr)^{\frac{3}{2}}\,. \label{eq:Ppair_sfE}
    \end{split}
\end{equation}
The general trend of the behavior of the pair probabilities~\eqref{eq:Ppair_sfE} with $d$ again resembles that of the associated polarization flip probabilities~\eqref{eq:Peperpe_strongB}.
Notably the $d=3$ result exhibits a distinctly different dependence on the probe photon polarization parameterized by $\beta$: While the leading strong-field terms in \Eqref{eq:Ppair_sfE} for $0\leq d\leq2$ vanish identically for $\beta\to0$, the $d=3$ contribution remains manifestly finite in this limit.

\section{Conclusions and Outlook}\label{sec:concls}

In this work we analyzed the impact of a weak localization of the background field on nonlinear quantum vacuum signals probed by photons traversing this field.
To this end we considered two different background field configurations, namely the case of a purely magnetic $\vec{B}$ or electric $\vec{E}$ field pointing in a fixed direction, and a crossed field fulfilling $\vec{B}\perp\vec{E}$ and $|\vec{B}|=|\vec{E}|$, where the directions of $\vec{B}$ and $\vec{E}$ are once again fixed.
The latter configuration can be considered as a toy-model of a linearly-polarized laser field that does not resolve the modulation of the field with the laser frequency.

Our study heavily relied on the possibility of explicit analytical insights into the Heisenberg-Euler effective action and the photon polarization tensor in a constant electromagnetic field at one loop.
As we pointed out in detail, aiming at analytical insights into weakly localized field configurations by using these results as starting point, the specific structure of their propertime representations suggests to focus on the study weakly localized field configurations with Lorentzian amplitude profile~\eqref{eq:calE}.
This allowed us to construct relatively compact expressions for the relevant one-loop photon polarization tensors in the presence of background field inhomogeneities characterized by a Lorentzian amplitude profile and featuring different numbers of inhomogeneous directions $0\leq d\leq3$.
The resulting expressions are accurate at leading order in a slowly-varying field approximation and their complexity is of the level of their constant-field analogues.

Our results for the magnetic/electric field case derived on the basis of the Heisenberg-Euler effective action allow for the analysis of generic probe photon scattering processes, but are by construction limited to the regime of low-energy photons.
On the other hand, those for the crossed field case obtained from the photon polarization tensor are restricted to the study of probe-photon forward scattering phenomena, but can be employed for arbitrarily large probe photon momenta.
In the present work, we studied both of these results in full detail. Aside from their all-order perturbative weak-field expansions, we mainly focused on parameter regimes in which analytical insights are possible and the coupling to the background field needs to be accounted for in a manifestly non-perturbative way.  
Our main interest was in the $d$ dependence of the scaling of the different scalar components constituting the respective expressions for the photon polarization tensor with the peak field strength of the Lorentzian amplitude profile~\eqref{eq:calE} in these non-perturbative parameter regimes.

As particular examples of physical signatures allowing to make the studied effects, at least in principle, accessible in experiment we discussed the vacuum-polarization-induced polarization-flip phenomenon experienced by linearly polarized probe photons traversing the external electromagnetic field, and their polarization-sensitive absorption in the background field, or equivalently, probe-photon induced electron-positron pair production.

Our considerations could be extended in various ways. 
We believe that a particularly interesting and prospective avenue for future research would be to adopt a similar approach towards the study of quantum vacuum processes at higher-loop order. 

\section*{Acknowledgments}

This work has been funded by the Deutsche Forschungsgemeinschaft (DFG) under Grant Nos. 392856280 and 416607684 within the Research Unit FOR2783/2.

\end{document}